\documentclass[12pt]{article}

\usepackage{graphicx}

\usepackage[paperheight=11.in, paperwidth=8.5in, left=0.75in, right=0.75in, top=0.5in, bottom=0.5in, nohead, nofoot, width=8.0in, height=10.00in]{geometry} 
\usepackage{amsmath}
\usepackage{rotating}
\usepackage{lscape}

\usepackage{amssymb}
\usepackage{gensymb}

\usepackage{natbib,twoopt}

\begin{document}




\title{Dynamics of Saturn's Great Storm of 2010-2011 from Cassini ISS and RPWS}

\begin{center}
\author{Kunio M. Sayanagi$^{1, 2}$,Ulyana A. Dyudina$^{2}$, Shawn P. Ewald$^{2}$,\\ Georg Fischer$^{3, 4}$, Andrew P. Ingersoll$^{2}$, William S. Kurth$^{4}$,\\ Gabriel D. Muro $^{2, 5\dagger}$, Carolyn C. Porco$^{6}$, and Robert A. West$^{7}$}
\end{center}


\baselineskip18pt


\maketitle 
\begin{noindent}
\\
\emph{$^{1}$Department of Atmospheric and Planetary Sciences, \\Hampton University, Hampton, VA, 23668, USA}\\
\emph{$^{2}$Division of Geological and Planetary Sciences, \\California Institute of Technology, Pasadena, CA, 91125, USA}\\
\emph{$^{3}$Space Research Institute, Austrian Academy of Sciences, \\A-8042 Graz, Austria}\\
\emph{$^{4}$Department of Physics and Astronomy, \\University of Iowa, Iowa City, Iowa, 52242, USA}\\
\emph{$^{5}$Department of Earth and Planetary Sciences, \\University of California Santa Cruz, Santa Cruz, CA, 95064, USA}\\
\emph{$^{6}$Cassini Imaging Central Laboratory for Operations, \\Space Science Institute, Boulder, CO, 80301, USA}\\
\emph{$^{7}$Jet Propulsion Laboratory, \\California Institute of Technology, Pasadena, CA, 91109, USA}\\
\emph{$^\dagger$Current Affiliation: Lunar and Planetary Lab, \\University of Arizona, Tucson, Arizona, 85721, USA}\\
\end{noindent}

\bigskip

\begin{center}
\scriptsize
Copyright \copyright\ 2012 Kunio M. Sayanagi and co-authors
\end{center}

\newcommand{\sep}{; }


\pagebreak

\begin{flushleft}
Number of pages (main text and figure captions): 44\\
Number of tables: 3\\
Number of figures: 17\\
\end{flushleft}

\noindent
\textbf{Proposed Running Head:}\\
  Dynamics of Saturn's Great Storm of 2010-2011

\vspace{3cm}
\noindent
\textbf{Please send Editorial Correspondence to:} \\
Kunio M. Sayanagi \\
  Department of Atmospheric and Planetary Sciences \\
  Hampton University, Hampton, Virginia, 23666 \\
  Email: \texttt{kunio.sayanagi@hamptonu.edu}\\

\vfill

\pagebreak

\noindent
\textbf{ABSTRACT: }
Saturn's quasi-periodic planet-encircling storms are the largest convecting cumulus outbursts in the Solar System. The last eruption was in 1990 \citep{Sanchez-Lavega_1994}. A new eruption started in December 2010 and presented the first-ever opportunity to observe such episodic storms from a spacecraft in orbit around Saturn \citep{Fischer_etal_2011Nature, Sanchez-Lavega_etal_2011Nature, Fletcher_etal_2011Science}. Here, we analyze images acquired with the Cassini Imaging Science Subsystem (ISS), which captured the storm's birth, evolution, and demise. In studying the end of the convective activity, we also analyze the Saturn Electrostatic Discharge (SED) signals detected by the Radio and Plasma Wave Science (RPWS) instrument. The storm's initial position coincided with that of a previously known feature called the \emph{String of Pearls} (SoPs) at 33\degree N planetocentric latitude. Intense cumulus convection at the westernmost point of the storm formed a particularly bright ``head'' that drifted at $-$26.9$\pm$0.8~m~s$^{-1}$ (negative denotes westward motion). On January~11, 2011, the dimensions of the head were 9200~km and up to 34,000~km in the north-south and east-west dimensions, respectively. RPWS measurements show that the longitudinal extent of the lightning source expanded with the storm's growth. The storm spawned the largest tropospheric vortex ever seen on Saturn. On January~11, 2011, the anticyclone was sized 11,000~km by 12,000~km in the north-south and east-west directions, respectively. The between January and September 2011, the vortex drifted at an average speed of $-$8.4~m~s$^{-1}$. We detect anticyclonic circulation in the new vortex. The vortex's size gradually decreased after its formation, and its central latitude shifted to the north. The storm's head moved westward and encountered the new anticyclone from the east in June 2011. After the head-vortex collision, the RPWS instrument detected that the SED activities became intermittent and declined over $\sim$40~days until the signals became undetectable in early August. In late August, the SED radio signals resurged for 9~days. The storm left a vast dark area between 32\degree N and 38\degree N latitudes, surrounded by a highly disturbed region that resembles the mid-latitudes of Jupiter. Using ISS images, we also made cloud-tracking wind measurements that reveal differences in the cloud-level zonal wind profiles before and after the storm.

\noindent
\textit{Keywords:}
SATURN, ATMOSPHERE; ATMOSPHERES, DYNAMICS; METEOROLOGY

\pagebreak


\section{Introduction\label{intro}}
Cumulus convecting storms on Saturn are known for their episodic behavior. Convective events on Saturn can be identified visually and through radio noise. The events occur intermittently within a small range of latitudes and are interspersed with quiescent periods that last for a year or longer \citep{Sromovsky_etal_1983, Dyudina_etal_2007, DelGenio_etal_2007}. The largest of these encircle the planet and are sometimes called Great White Spots \citep{Sanchez-Lavega_1994}. These giant storms exhibit a $\sim$30-year periodicity ($\sim$1~Saturn Year), and have been recorded in 1876, 1903, 1933, 1960 and 1990, in which the storms lasted for 26, 150, 44, 41, and 55 days, respectively \citep{Alexander_1962, Sanchez-Lavega_1994}. All documented planet-encircling storms have occurred in the northern hemisphere, alternately erupting in mid-latitude and equatorial regions \citep{Alexander_1962, Sanchez-Lavega_1994}. These storms exhibit some morphological similarities to Jupiter's South Equatorial Belt (SEB) revival events \citep{Sanchez-Lavega_Gomez_1996}; however, their convective areas occupy an area much narrower than the surrounding jets, and move at speeds that reflect the background zonal mean wind. An SEB revival event may last up to 7 months; however, unlike Saturn's planet-encircling storms, its relatively small convective cores do not form a large organized structure.

The 1990 equatorial planet-encircling storm of Saturn was an early scientific target of the Hubble Space Telescope \citep{Sanchez-Lavega_etal_1991, Sanchez-Lavega_1994, Beebe_etal_1992, Westphal_etal_1992}. The data collected during the 1990 storm enabled examinations of Saturn's equatorial dynamics. Based on comparing the equatorial wind speeds between Voyager and Cassini measurements, \cite{Perez-Hoyos_Sanchez-Lavega_2006} suggested that the 1990 storm caused a deceleration in the equatorial jet. Numerical simulations by \cite{Sayanagi_Showman_2007} tested the hypothesis that the storm caused a deceleration. They demonstrated that the observed apparent wind change can be explained by the deceleration of the wind caused by the storm in combination with the change in the altitudes of the tracked clouds. This interpretation is consistent with the radiative transfer analysis of \cite{Perez-Hoyos_Sanchez-Lavega_2006}.

A new planet-encircling storm started on December~5, 2010. The 2010-2011 outburst was the first planet-encircling storm to be studied from a spacecraft in orbit around Saturn. The Cassini orbiter's Radio and Plasma Wave Science (RPWS) instrument detected radio pulses emitted by lightning discharges in the storm \citep{Fischer_etal_2011Nature}, and the Composite Infrared Spectrometer (CIRS) measured the storm's large-scale heating of the stratosphere and a longitudinal thermal perturbation that is consistent with the formation of a new anticyclonic vortex \citep{Fletcher_etal_2011Science, Fletcher_etal_2012}. Also, ground-based observations captured the development of the large scale disturbance \citep{Sanchez-Lavega_etal_2011Nature}. The mechanism that controls the intermittent appearances and scales of these extreme outbursts is unknown; a better understanding of the conditions that precede these episodes is one step toward unveiling the storms' origin.

Here, we analyze Cassini ISS images and RPWS data to investigate the temporal evolution of the 2010-2011 planet-encircling storm. Our observations cover the temporal evolution of the planetary-scale convective disturbance from its beginning on December~5, 2010 until the end of 2011. The rest of our report is organized as follows. Section~2 describes the ISS image sets and the processing applied to them in our study. Section~3 presents the chronological development of the 2010-2011 storm. Section~4 summarizes our findings.

\section{Cassini Data Sets}
This report presents images captured by Cassini's wide-angle camera at 60-120~km per pixel image scale. Our study analyzes images acquired using the wide-angle camera's CB2 (750~nm), MT2 (727~nm), and MT3 (889~nm) filters to study the vertical cloud structure. We follow the standard image processing procedure previously used by others to process ISS images of Saturn \citep{Porco_etal_2005, Vasavada_etal_2006, DelGenio_etal_2007, West_etal_2010cisscal}. We use the camera geometric model and the photometric calibration software CISSCAL version~3.4 described by \cite{Porco_etal_2004} and released on April~9, 2006. The images are navigated by fitting the planet's limb using the spacecraft and planetary ephemeris provided by the Jet Propulsion Laboratory (JPL)'s Navigation and Ancillary Information Facility.

The calibrated images are then photometrically flattened to remove the contrast caused by variation in the solar illumination angle. The illumination effects are compensated by dividing the calibrated disk image by the Minnaert scattering correction factor,
\begin{equation}
f = \textrm{cos}^{k}(i) \textrm{cos}^{k-1}(e),
\end{equation} 
where $i$ is the incidence angle and $e$ is the emission angle. We used $k$=0.5 for our corrections. 

We map-project the photometrically flattened images into a cylindrical latitude-longitude projection. In projecting the images, we assume that Saturn's equatorial and polar radii are 60268~km and 54364~km, respectively. Projected maps have 0.1$\times$0.1~degree per pixel in latitude$\times$longitude. We use planetocentric latitude and Eastward System III Longitude (i.e., reference frame of $\Omega=$1.638$\times10^{-4}$~s$^{-1}$, \cite{Seidelmann_etal_2007}) for the rest of this report. To further enhance the contrast, we apply a high-pass spatial filter to the map-projected images to remove the residual large-scale variation caused by scattering in the stratospheric haze. We remove variation greater than 20\degree~in both latitude and longitude. Our high-pass filtering may remove large-scale cloud structure; we justify the use of spatial filtering because important small-to-medium scale features are difficult to resolve unless large scale background contrast variations are removed. Also, any large scale, smooth structure that may exist would be difficult to meaningfully distinguish from the contrast variation caused by the stratospheric haze. We then assembled the contrast-enhanced images into mosaics. Each mosaic contains images that are acquired within one Saturnian rotation to provide a full longitudinal coverage. 

In addition to studying the morphology and dynamics of the new planet-encircling storm, we also measure cloud motions to study the atmospheric dynamics of the storm region. In this study, we use two methods to analyze the cloud motions. First, we adopt TRACKER3, an automatic feature tracker developed at JPL, to perform Correlation Imaging Velocimetry (CIV) wind measurements. TRACKER3 was previously applied to analyze wind fields on Jupiter and Saturn \citep{Salyk_Ingersoll_etal_2006, Vasavada_etal_2006}; here, we use TRACKER3 to measure the motions in and around the new anticyclonic vortex that was spawned by the storm. Second, to measure the temporal evolution of the zonal wind profile, we apply the one-dimensional correlation method (1Dcorr); the method was previously used by \citet{Limaye_1986} and \cite{Garcia-Melendo_Sanchez-Lavega_2001} for Jupiter. CIV is able to extract two-dimensional wind fields while 1Dcorr can only determine dominant zonal motions; on the other hand, CIV requires high-contrast images while 1Dcorr can return reliable measurements from relatively low-quality images. Not all of our images had enough contrast to be suitable for CIV applications; thus we used CIV for measuring the wind field around the vortex, while we used 1Dcorr to detect changes in the zonal jet speeds.

We also present analysis of RPWS data that shows near-simultaneous coverage of the end of the convecting phase of the storm. The data analysis method is described by \cite{Fischer_etal_2006Icarus}.

\section{Temporal Evolution of the 2010-2011 Planet-Encircling Storm}
In this section, we describe the development of the storm in a chronological order. Figure~1 shows the temporal evolution of Saturn's full-disk appearance in real color over the course of the storm. The new storm was first captured by chance on December~5, 2010, and is barely visible in Figure~1 as a faint spot in the northern hemisphere near the terminator (color-enhanced view highlights its presence, to be presented in Fig.~4). The storm subsequently affected the entire latitude circle around 33\degree N planetocentric latitude.

\subsection{String of Pearls: a Precursor to the Storm?}
Before the new storm, a feature known as the String of Pearls (SoPs) was previously detected using the Cassini Visible and Infrared Mapping Spectrometer (VIMS) \citep{Momary_etal_2006, Choi_Showman_Brown_2009}. Figure~2 presents an ISS image of the SoPs. The image was acquired on March~29, 2008 using the CB2 filter, in which the ``pearls'' can be seen as a chain of 17~dark, featureless spots that occupy latitudes between 32.5\degree N and 34.5\degree N. In VIMS 5-micron images, the SoPs appeared as bright spots, which were interpreted to be holes in the cloud deck from which thermal radiation escapes from depth. SoPs are visible in the visible bands (RGB) and CB3 filter albeit with lower contrast, while they are completely indecipherable in MT2 and MT3, suggesting that the structure does not extend above tropopause. Our detection of the SoPs in the CB2 band as dark featureless areas is consistent with this previous interpretation. \cite{Sayanagi_etal_2012sops} showed that the individual \emph{pearls} have cyclonic vorticity, and the feature's longitudinal drift can be fitted by a constant speed of $-$22.42$\pm$0.2~m$^{~}$s$^{-1}$ (i.e., $-$2.28\degree~per Earth day in System III) between September~2006 and July~2010. For the rest of the report, eastward (westward) speeds is denoted by positive (negative) values.

Figure~3a demonstrates that the new storm erupted from the SoPs. The figure presents the motion of the SoPs from September~2006 to December~2010. When the storm started on December~5, 2010, it was centered at 114.1\degree~east longitude and 32.6\degree N latitude. On December~5, the SoPs was predicted to occupy longitudes between 62\degree - 161\degree~(the December 5th image's poor image contrast does not allow SoPs detection). Thus, the storm's initial latitude-longitude coordinates coincided with those of the SoPs. The SoPs was last detected in an ISS image on December~23, 2010; since then, the storm has engulfed the entire latitude zone and the SoPs has not been seen.

\subsection{Storm's initial evolution}
Figure~4 shows the temporal evolution of the storm-affected region. The figure presents latitude-longitude projected, composite-color mosaics in chronological order. In Fig.~4, the colors convey the cloud-top heights. The red, green and blue color channels of the mosaic are assigned, respectively, to images acquired in a continuum band (CB2), a moderate methane absorption band (MT2), and a strong methane absorption band (MT3). Because methane is well-mixed, features that appear bright in the MT2 and MT3 absorption bands must reside at high altitudes \citep{Tomasko_etal_1984}. Consequently, in this color scheme, the altitude of detected features generally increases in order of red-green-blue -- white indicates that light is scattered in all three wavelengths bands by aerosols at high altitudes. Quantitative determination of cloud altitudes require extensive radiative transfer modeling, which is beyond the scope of our study. 

Prior to the storm, the region that spawned the storm was characterized by a gap in the high-altitude haze around 31.5\degree N latitude that appears as a deep red band in Fig.~4 panel 2008-03-29, and wispy tropospheric clouds (which appear red in Fig.~4) that had typical length scale of $\sim$500~km. Note that, even though the SoPs is present within the field of view of Fig.~4's panel 2008-03-29 and 2010-12-23, the feature cannot be seen because it is does not leave any decipherable signature in MT2 and MT3. 

Cassini ISS captured the new storm's onset on December~5, 2010 (Fig.~1 panel 2010-12-05), the same day the RPWS started detecting lightning originating from the longitude of the ISS-detected storm \citep{Fischer_etal_2011Nature}. The onset of the storm was also captured by ground-based observers \citep{Sanchez-Lavega_etal_2011Nature}. The storm appears as a red spot in our false-color scheme near the eastern end of the image, indicating that it was a deep cloud at this time. On December~5, the storm covered an area that extended 1920$\pm$130~km in east-west and 1300$\pm$130~km in north-south dimensions, and had an area of approximately 1.5$\times$10${^6}$~km${^2}$. The storm cloud was centered at 115.4$\pm$0.1\degree~longitude and 32.6$\pm$0.1\degree N latitude. The new spot exhibited a sharp outline and covers a larger area than a regular cloud in this latitude zone, as shown in panel 2008-03-29. Outside of the area covered by the storm clouds, the general cloud morphology appears unchanged, where the gap in the high-altitude haze persists and clouds with $\sim$500-km horizontal scale (unchanged since before the storm) remain prevalent.

\subsection{Anatomy of the Storm}
Figure~5 presents a full-longitude mosaic of the storm on January~11, 2011. The figure illustrates that the storm's cloud body consisted of three parts. We call the westernmost bright cloud the storm's \emph{head}. Bright clouds billow to the east of the head out to the large anticyclonic vortex (AV). To the east of the AV, clouds acquire a turbulent appearance without well-defined edges -- we call this part the \emph{tail} of the storm. Here, the definitions of the ``head'' and ``tail'' follow those of \cite{Fletcher_etal_2011Science}, \cite{Sanchez-Lavega_etal_2011Nature} and \cite{Sanchez-Lavega_etal_2012gws}.

Figure~3b shows the longitudinal motions of the head and AV. The longitudes of the head's western leading edge throughout its lifetime are listed in Table~1. The best-fit function that expresses the eastward System III longitude of the head $\theta_\textrm{head}$ as a function of Julian day $t_\textrm{Julian}$ is
\begin{equation}\label{e:head_motion}
\theta_\textrm{head}(t_\textrm{Julian}) = -2.79 (t_\textrm{Julian} -2455000) +1611.
\end{equation}

This fit is indistinguishable from that reported by \cite{Sanchez-Lavega_etal_2012gws}. Between December~5, 2010 and the head's last sighting on June~14, 2011, its best-fit propagation rate is $-$2.79$\pm$0.08 \degree~day$^{-1}$ ($-$26.9$\pm$0.8~m~s$^{-1}$). During this period, the drift rate fluctuated between $-$2.6\degree~day$^{-1}$ and $-$3.2\degree~day$^{-1}$. The image navigation uncertainty (0.2\degree~in both latitude and longitude for the worst cases) was negligible for the purpose of calculating the head's drift rate. The head's average drift rate is 0.5\degree~day$^{-1}$ faster than that of the SoPs; consequently, the head left the region that was predicted to be occupied by the SoPs in mid-April 2011 (shown in Fig.~3a). The clouds to the west of the head maintained the pre-storm morphology until the tail completed its circumpropagation of the latitude zone. The head appears bright in all of CB2, MT2 and MT3 images and thus appears white in Fig.~4. The sharp edge of the head is accompanied by a dark outlining region immediately outside of the head as can be seen in Fig.~5. This darkening is most pronounced in MT2 and MT3, and the geometry of the darkening is inconsistent with any possible shadows cast by the storm clouds on surrounding cloud deck at a lower altitude. The dark region may indicate the descending circulation that surrounds the intense updraft of the cumulous storm.

\subsection{Formation of a Large Anticyclonic Vortex}
The AV formed between December~5 and 24. On December~24, (Fig.~4 panel 2010-12-24), the AV had a roughly circular core that appeared particularly bright in MT3 (blue in Fig.~4 and 5). The core had a diameter of approximately 4250~km and was centered at 76\degree~longitude and 33\degree N latitude. By January~11, the AV measured 11,000~km by 12,000~km in the north-south and east-west dimensions (Fig~4 panel 2011-01-11 and Fig.~5), a size that rivaled Jupiter's Oval BA \citep{Simon-Miller_etal_2006_OvalBA}. The AV's blue hue in Fig.~4 and 5 indicates clouds that reside at a higher altitude than the surrounding clouds that appear white. \cite{Sanchez-Lavega_etal_2011Nature} determined that the top of the clouds in the storm's head, which appear white in Fig.~4, reached $\sim$150~mbar, and we interpret that the core of the AV extended to a higher altitude.

Earth-based observations measured the size of the dark oval that seems to correspond to the core of the AV. Infrared mages captured on January~19 at the VLT showed the dark oval's meridional$\times$zonal size to be 5500$\times$4000~km \citep{Fletcher_etal_2011Science}, while Hubble Space Telescope observation in visible wavelengths on March~ 12 showed the size to be 6000$\times$7800~km \citep{Sanchez-Lavega_etal_2012gws}. 

Figure~6a presents wind vectors within the AV on January~11, 2011, demonstrating that the vortex has an anticyclonic circulation. We deduced the wind field in the vortex using TRACKER3. Supplementary Movie~1 also shows the cloud motion within the AV. Similar to large anticyclones on Jupiter \citep{Dowling_Ingersoll_1988, Simon-Miller_etal_2006_OvalBA, Choi_etal_2007, Cheng_etal_2008, Hueso_etal_2009, Asay-Davis_etal_2009, Sussman_etal_2010}, highest wind speeds within the AV, which exceeded 120~m~s$^{-1}$, are detected in an outer annular region. For these images, the limb of the planet could not be used to navigate the images, thus the pointing information of the spacecraft was used to navigate the images. The worst-case uncertainty in the wind magnitude arising from image navigation is $\pm$16.5~m~s$^{-1}$.

Figure~6b shows the relative vorticity derived from the wind field presented in Fig.~6a, showing that the relative vorticity reached $-$6$\times$10$^{-5}$~s$^{-1}$ in the vortex's annular region. To fill missing vectors and alleviate the effects of spurious measurements, the wind field was smoothed over 6 vector grids (2.3\degree~in both latitude and longitude) before calculating the curl of the field. Note that the image navigation uncertainty can only introduce uniform offset in the wind vectors, and does not severely impact the vorticity measurements. We estimate that the uncertainty in the relative vorticity values that have been masked by the smoothing is $\pm$1$\times$10$^{-5}$~s$^{-1}$.

Between December 24, 2010 and June~14, 2011, the AV maintained a mean drift speed of $-$0.85\degree~day$^{-1}$ ($-$8.4~m$^{~}$s$^{-1}$); Table~2 chronologically lists the longitudes of the center of the AV. The best-fit function that expresses its eastward System III longitude $\theta_\textrm{AV}$ as a function of Julian day can be written as
\begin{equation}\label{e:AV_motion}
\theta_\textrm{AV}(t_\textrm{Julian}) = -0.85 (t_\textrm{Julian} -2455000) + 540.
\end{equation}
During this time, the drift rate fluctuated between $-$0.6\degree~day$^{-1}$ and $-$1.4\degree~day$^{-1}$. The AV motion deviated from this linear fit after September~2011 as will be noted later.

Figure~7 presents the cross-sectional wind profiles of the AV on January~11, 2011. The north-south cross-section of the eastward wind component through the AV's central longitude (54\degree) is shown in Fig.~7a, demonstrating that the tangential wind speed reached $+$100~m~s$^{-1}$ at the north side of the AV's annular region. The wind speed at the south side reached $-$60~m~s$^{-1}$. The tangential wind speed within the vortex clearly exceeds that of the background zonal mean wind, also shown in Fig.~7a. Figure~7b shows the east-west cross-section of the northward wind component through the AV's central latitude (35.5\degree), demonstrating that the northward speed at the western edge of the AV exceeded $+$50~m~s$^{-1}$ while it reached $-$100~m~s$^{-1}$ at the eastern edge.

These characteristics are similar to those of Jupiter's Oval BA, an anticyclone of similar size \citep{Simon-Miller_etal_2006_OvalBA, Cheng_etal_2008, Asay-Davis_etal_2009, Hueso_etal_2009, Sussman_etal_2010, Choi_etal_2010}. Like Oval BA, in the current AV measurement, the magnitude of the eastward tangential velocity at the poleward edge of the anticyclone is faster than the westward tangential velocity at the equatorward edge. Also, the magnitude of the equatorward tangential velocity at the eastern edge of both anticyclones is faster than the poleward tangential velocity found at their respective western edge. The magnitudes of the tangential velocities are generally similar between Oval BA and the AV. We note that the background zonal wind has a stronger shear for the AV than for Oval BA. On both Jupiter and Saturn, the cyclonic shear zone at the equatorward side of the vortex harbors a stronger shear than in the anticyclonic shear zone to the poleward side. In the cyclonic side, Saturn's zonal mean wind in the vicinity of the AV changes from $+$60~m~s$^{-1}$ to $-20$~m~s$^{-1}$ between 30\degree N to 34\degree N, averaging 1.7$\times$10$^{-5}$~s$^{-1}$. This compares to the cyclonic region at the equatorward side of Oval BA where the wind changes from $+$50~m~s$^{-1}$ at 24\degree S to $-$10~m~s$^{-1}$ at 29\degree S, averaging 1.0$\times$10$^{-5}$~s$^{-1}$.

Long-lived vortices are less common on Saturn than on Jupiter, and not many Saturnian vortices have been studied in detail. \cite{Garcia-Melendo_Sanchez-Lavega_Hueso_2007} and \cite{DelRio-Gaztelurrutia_etal_2010} are two notable studies of Saturnian vortices to date. The former studied the properties of a long-lived anticyclonic vortex, called the \emph{Brown Spot} (BS) \citep{Sromovsky_etal_1983}, that seems to have persisted since the Voyager era in 1980-81 until the arrival of Cassini; and the latter focused on a long-lived cyclonic vortex in the southern hemisphere. Both vortices are characterized by tangential wind velocities exceeding the ambient wind field by $\sim$10-20~m~s$^{-1}$, and their horizontal length scales are smaller than a single shear-zone on one side of the jet each of the vortices resides in. From these viewpoints, Jupiter's Oval BA and Saturn's new AV belong to a different class of vortices that harbor winds much stronger than the background zonal jets.

\subsection{Storm's Growth}
Due to the difference in the drift speeds of the head and AV, their spacing widened at an approximately constant rate of 2.0\degree~day$^{-1}$ as shown in Figs.~3b and 4. As the space between the head and the AV widened, the storm's bright cloud that occupied the region between the two features covered 1.75$\times$10$^8$~km$^2$ on December~24, 2010, 2.82$\times$10$^8$~km$^2$ on January~2, 2011, and 4.62$\times$10$^8$~km$^2$ on January~11, 2011. Since the beginning of the storm on December~5 and until January~11, we measured an approximately constant expansion rate of 140~km$^2$s$^{-1}$. Using ground-based telescopes, \cite{Sanchez-Lavega_etal_2011Nature} measured a peak expansion rate of 212~km$^2$s$^{-1}$ during a period in December~14-16 when Cassini was not taking data. Radiative transfer calculations by \cite{Sanchez-Lavega_etal_2011Nature} placed the storm's cloud top at 150~mbar. If we assume that the cloud layer represented a mass detrainment layer that occupied one scale-height between 150-400~mbar and the detrainment is continuous, we are able to calculate the storm's mass flow rate to be 3.8$\times$10$^{11}$~kg~s$^{-1}$. After January~11, the cloud boundaries of the storm's body became fuzzy and precise determinations of the cloud coverage became difficult. The storm maintained the expansion rate of $\sim$200~km$^2$s$^{-1}$ over this entire 6~month period, assuming an area based on the 300,000~km circumference of the planet and the 10,000 km width in latitude.

As the storm grew in size, the ``encircling'' of the planet happened in multiple stages. Here, by \emph{encircling}, we mean an encounter between the head of the storm and the other discrete features of the storm. All features move westward but at different speeds; because the head is the fastest-moving feature, it overtook the other slower-moving features one by one. First, the head met the turbulent tail by late January 2011. The head encountered the bright cloud material that emanated from the storm before March. Finally, the head encountered the AV in mid-June; as the main body of the storm cloud was bound between the head and the AV, the storm cloud completely engirdled the latitude zone when the head met the AV.

\subsection{Dark Ovals}
In mid-January, a train of dark ovals (DOs) gradually formed in the storm's tail. The ovals are centered at 25\degree N latitude. In Figs.~4 and 5, the DOs have a deep blue color because they are dark in CB2 and MT2 images and bright in MT3; the appearance suggests that the DOs have an optically thin scatterer in the stratosphere that is sensed in the MT3 and no scattering layer underneath. The DOs and the SoPs are distinct features with completely different dynamical and morphological characteristics. The SoPs drifted at $-$22.42~m~s$^{-1}$ (i.e., westward), whereas the DOs drifted eastward at $+$65~m~s$^{-1}$. The SoPs resided at 33\degree N while the DOs formed at 25\degree N. The SoPs appeared dark in CB2 and left no discernible signature on MT2 or MT3; in contrast, the DOs were dark in CB2 and MT2, and bright in MT3. All DOs disappeared by April~25 (panel 2011-04-25) after an observational gap of one month. Table~3 lists the longitudes of the detected DOs. We detected DOs in images acquired between January~12 and March~21. We detected up to 6~DOs simultaneously; however, note that not all data covers a full 360\degree~of longitudes, thus detecting fewer DOs at a given time does not necessarily mean fewer DOs were present. We were unable to detect the vorticity of DOs. Similar dark spots were also detected in previous smaller SED storms in the southern hemisphere \citep{Dyudina_etal_2007}. \cite{Baines_etal_2009lightning} proposed that the dark color left after a lightning storm is a signature of elemental carbon and condensates upwelled from deeper levels; whether this hypothesis applies to the DOs or not is unclear. The elemental carbon hypothesis does not explain how the DOs acquire the elemental carbon because they do not emerge from the convecting region. We also note that all dark spots previously noted on Saturn including the ``Brown Spot'' \citep{Smith_etal_1982science, Sromovsky_etal_1983}, and the dark vortices in the southern hemisphere \citep{Vasavada_etal_2006, Dyudina_etal_2007} were located in anticyclonic shear zones; in contrast, the present DOs formed in a cyclonic shear zone.

\subsection{Head-AV Collision and End of the Convective Storm}
In late June 2011, the head of the storm caught up with the AV after the vortex's position fell 360\degree~longitude behind (to the east) of the head due to the AV's slower westward drift. Figure~4 panel 2011-06-14 shows the head and the AV 10~days before their positions were predicted to intersect on June~24. Ground-based observation by \cite{Sanchez-Lavega_etal_2012gws} show that the head and the AV made contact around June~15, and the bright cloud materials of the head first disintegrated into three parts before disappearing between June~14 and June~19. On July~12 (Fig~4 panel 2011-07-12), the ISS imaged the storm for the first time after the predicted collision between the head and the AV. Our images reveal that, after the collision, the head is no longer visible while the AV is still present. Due to the one-month-long gap in our observational coverage around the time of head-AV collision in mid-June, 2011, we are unable to be absolutely certain that the vortex that emerged out of the collision is the continuation of the original AV; however, since the post-collision AV's longitudinal position is consistent with the pre-collision AV, we conclude that a single vortex persisted throughout our observation record.

The collision of the head and AV marked a major change in the storm's convective activity and SED signals. Each SED is thought to come from a single lightning strike. The propagation of SED signals through the atmosphere and its detection by RPWS is described by \cite{Zarka_etal_2008}. The SED signals are detectable by the RPWS while an active storm, which continuously emits SED signals, is on the spacecraft-facing side of the planet as illustrated in Fig.~8a, showing the geometry of the observation on December~7, 2010 as an example. The signal detection starts when the SED source rises above the western horizon, and ends when it sets to the eastern horizon. Until the end of June, the SED signals stayed detectable until the storm's head was $\sim$30\degree~beyond the visible horizon, which is believed to be due to the so-called ``over-horizon'' effect \citep{Zarka_etal_2008} caused by temporary trapping of the SED radio waves below Saturn's ionosphere, which is depicted in Fig.~8c. Around the periapses, the SED longitudes are shifted to the east. The shift happens during a short period when Cassini is on the morning side of Saturn as depicted in Fig.~8d. This causes the over-horizon effect at the beginning of an SED episode before the source rises over the visible horizon; for the rest of the time, when Cassini is on the afternoon side (shown in Fig.~8a-c), the over-horizon effect occurs after the source sets below the eastern horizon.

Figure~9 shows the longitudinal drift of the SED source measured by the RPWS instrument during the storm. In Fig.~9, the rise of the SED signal source over the western visual horizon is denoted by the bottom of each black vertical line, and the setting of the source over the eastern horizon is marked by the top of each line. Figure~9 demonstrates that the SED's source had a western edge coinciding the head of the storm, and its signals were continuous between the storm's beginning on December~5, 2010 and the head-AV collision in June~2011. Figure~9 can also be used to estimate the longitudinal extent of the SED source. In the figure, the longitudinal range bounded by the blue lines denote the longitudinal range facing the Cassini spacecraft when the storm head passes the sub-spacecraft longitude. When the SED signals are marked below the lower blue line, we interpret that they originated in regions to the east of the head. Figure~9 shows that the SED source region steadily grew eastward after December~2010. Its longitudinal extent reached approximately $\sim$90\degree~in March~2011, and $\sim$180\degree~in May~2011. Figure~8b illustrates the observation geometry of March~2011. When the longitudinal extent of the source exceeds 180\degree, RPWS detects the SED signals continuously because a portion of the source is always within a line of sight; consequently, we are unable to estimate the source size when it spans more than 180\degree~in longitude (this assumes that the radio horizon coincides the visual horizon; when the over-horizon effect places the radio horizon behind the visual horizon, RPWS can receive continuous SED signals when the source region extends less than 180\degree~in longitude). The SED source size exceeded 180\degree~longitude until late June, approximately coinciding the head's collision with the AV. 

Figure~10 shows that the SED signals recorded by the RPWS instrument became intermittent after the head-AV collision and the head's demise. Figure~10a shows the number of SED flashes per hour recorded by the RPWS as a function of time between mid-June until the end of August. As weaker SED flashes are not detected by the RPWS while the spacecraft is further away from Saturn, we use the spacecraft's distance from Saturn (shown in Fig.~10b) to normalize SED flash rate (normalized to 50~Saturn radii) and show the result in Fig.~10c. It is interesting to note that, even though the head-AV collision was a localized event, the SED flash rate decreased after the collision throughout the longitudes.

Figure~11 presents a full-longitude CB2-MT2-MT3 mosaic of the storm-affected latitudes on July~12, after the demise of the storm's head and during the declining of the SED signals. Figure~9 shows that, around July~12, the SED signal was detectable while the subspacecraft longitude was between 260\degree - 130\degree~longitude (i.e., spanning 230\degree of longitude). Assuming that the radio horizons roughly coincide the visual horizons, we estimate that the source had an extended longitudinal size between 350\degree - 40\degree~longitude (i.e., spanning 50\degree of longitude). Figure~11 shows a streak of bright cloud between 20\degree~and 40\degree~longitudes at 39\degree~latitude; we identify this cloud as a portion of the source of the SED signals during this time. In Fig.~11, the longitude of the storm head predicted by Eq.~\ref{e:head_motion} is marked with a diamond at 223\degree~E, showing that the SED source was 200\degree~away from the expected longitude of the storm's head if it had stayed active. The longitude of the AV predicted by Eq.~\ref{e:AV_motion} is marked with an asterisk; on July~12, the AV was slightly to the west of the predicted longitude and its drift motion had not deviated substantially from the constant-speed fit. 

Figures~9 and 10 also show a resurgence in the SED signals that started on August~19, which continued for 9~days. The SED signals of the August resurgence were detectable while the sub-spacecraft longitude is between $\sim$50\degree - 250\degree. The signal is consistent with a compact source centered around 130\degree~longitude, slightly to the east of the longitude of the storm head predicted by Eq.~\ref{e:head_motion}. Figure~9 shows the shift in the center longitude of the August resurgence. Figure~12 shows a full-longitude mosaic of CB2 images on August 24, 2011, in which we identify two bright clouds that are likely the source of the August SED resurgence, located at 125\degree~and 135\degree~longitude. After the August resurgence, convective activity was absent, and RPWS detected two weak SED activities for the remainder of 2011 (September~30 to October~6, and December~26-28). Between April 6 - 9, 2012, another weak SED activities were detected.

\subsection{Post-storm Cloud Morphology}
Figure~13 compares the cloud morphology in the visible wavelengths before (April~17, 2008, panel 2008-04-17) and after (August~11, 2011, panel 2011-08-11) the storm in enhanced RGB color. The region had been quiescent since Cassini's arrival at Saturn in mid-2004 and the SoPs was the only notable feature. The cloud maps show that, after the convective activity ceased, a vast dark region spanning up to $\sim$180\degree~in longitude appeared in the storm-affected latitudes between 31\degree N and 38\degree N. The region appeared dark in CB2 and visible continuum wavelengths and had a relatively bright appearance in MT2 and MT3 as shown in Fig.~11 captured on July 12, when the dark region spanned between 90\degree - 180\degree. Figure~12 presents the CB2-only view, in which the dark region extends between 350\degree - 140\degree~longitude, in which the darkness of the region is particularly notable; in CB2, the post-storm dark region represents the lowest-albedo region on Saturn after the storm. This appearance can be consistent with two scenarios; the first scenario has dark clouds around the 400-mbar level underlying a high-altitude haze \citep{Baines_etal_2011DPS}, and the second scenario has no deep clouds (i.e., clear between the 3-bar and 400-mbar levels) and a high-altitude haze. The second scenario would require an unknown absorber in the 750~nm band, as otherwise a cloudless region should appear bright from space due to the Rayleigh scattering by hydrogen molecules. The second scenario is analogous to the 5-micron hotspots of Jupiter. Galileo entry probe demonstrated that the region was clear \citep{Ragent_etal_1996}; however, viewed in visible wavelengths from orbit and Earth, the hotspots appear dark \citep{Simon-Miller_etal_2001}. \cite{Banfield_etal_1998} proposed methane absorption to explain the dark appearance; however, the adopted absorption profile was not consistent with the methane spectrum derived by \cite{Karkoschka_Tomasko_2010}. Another scenario that may explain the post-storm dark region is that chromophore particles are mixed to deep levels and they are absorbing the light. 

These scenarios should be distinguishable with VIMS 5-micron observations and/or Cassini Microwave Radiometer observations. If the ``deep'' cloud deck is indeed absent, the region should appear bright in both VIMS 5-micron channel and Cassini Microwave Radiometer maps, while the presence of an optically thick cloud layer would make the region appear dark to those instruments.

Figure~12 and Fig.~13 panel 2011-08-11 show that the post-storm dark region is surrounded by turbulent clouds whose spatial scales are substantially greater than those of the pre-storm clouds. The post-storm morphology exhibits a qualitative resemblance to that in the mid-latitudes of Jupiter. Our observations do not detect the SoPs in the aftermath of the storm; this may mean that the feature is obscured by overlying clouds, the cloud morphology is no longer responding to the dynamics of the SoPs, or the feature no longer exists. We also note that the AV persisted at least 4~months after it collided with the head.

\subsection{Change in Zonal Wind Profile}
Our measurements show that zonal wind profile after the end of the convecting phase is different from before and during the storm. Figures~14 and 15 present zonal wind profiles measured using the 1Dcorr method applied to images acquired using the CB2 filter, which senses the top of tropospheric cloud deck. 1Dcorr can be sensitive to wave propagation in addition to wind motions; to address this issue, we present Fig.~14 to illustrate that the zonal correlation distribution has a unique peak at every latitude covered in the measurement, and not likely to be confused by wave propagation. Figure~15 compares the maximum correlation zonal wind profiles (i.e., the zonal motion that maximizes the correlation values) before, during and after the storm. The figure demonstrates that measurements for May~7, 2008, September~7, 2010 and January~11, 2011 return zonal wind profiles that are essentially the same over much of the northern hemisphere except at the storm-perturbed latitudes. Figure~15b also shows that the head of the storm's $-$26.9~m~s$^{-1}$ propagation speed was substantially faster than the zonal mean wind speed at its latitude 32.6\degree N. In comparison, the zonal mean wind speed at the latitude was $-$15$\pm$2.1~m~s$^{-1}$ on May~7, 2008, $-$12$\pm$3.8~m~s$^{-1}$ on September~7, 2010, $-$6$\pm$1.7~m~s$^{-1}$ on January~11, 2011, and $-$22$\pm$4.2~m~s$^{-1}$ on August~5, 2011 (note that, on August~5, the storm head was no longer active). 

As shown in Fig.~15, the measurement made after the end of the convective phase on August~5, 2011 reveals wind speed changes. The latitudes to the south of 34\degree N exhibit a deceleration of the zonal wind profile by $\sim$30~m$^{~}$s$^{-1}$ while the profile to the north indicate an increase by $\sim$35~m$^{~}$s$^{-1}$. These changes in the zonal wind profile are greater than both the uncertainty arising from the image navigation and the width of the correlation peaks shown in Fig.~14, thus the wind speed changes detected in our measurements are greater than the uncertainties in our measurements. Interestingly, the zonal wind measurements during the Voyager flybys in 1980-81 by \cite{Sanchez-Lavega_etal_2000} indicate wind speeds between our pre- and post-storm measurements.

\cite{Fletcher_etal_2011Science} used Cassini CIRS temperature measurements to estimate the change in the vertical shear between September 2010 and January 2011 using the thermal wind equation, and detected wind speed changes greater than 40~m~s$^{-1}$ above the 50-mbar level. The wind change they derived is opposite of our measurements; they detected wind speed decrease to the north, and increase to the south of the storm latitude. Nevertheless, they detect little change between 1~bar and 100~mbar, and their measurements do not contradict our measurements. \cite{Achterberg_etal_2012DPS} and \cite{Fletcher_etal_2012} presented CIRS data that shows that, at the latitude of the storm, tropospheric temperature increased between January and August of 2011. Achterberg et al's measurements indicate $dT/dy > 0$ equatorward of the storm and $dT/dy <0$ poleward of the storm ($T$ is temperature and $y$ is northward distance). The temperature gradient suggests that, the wind speed above the 1-bar level decreased to the south of the storm and increased to the north. Thus, the CIRS vertical shear measurements support our interpretation that the change in the cloud-tracked winds represents a real wind speed change.

Our wind speed measurements do not rule out a possibility that the change in the cloud motion was caused by change in the altitude of the clouds. CIRS temperature analyses by \cite{Read_etal_2009} show that the zonal mean wind speed increases with height around 30\degree N while it decreases with height around 40\degree N. \cite{Garcia-Melendo_etal_2011saturn} also detects slight decay in the zonal wind with height at 40\degree N. The wind speed change detected in our zonal wind measurements is consistent with a decrease in the cloud altitudes in the regions both to the north and south of the storm; however, this would be surprising as the cumulus storm should have raised the cloud altitudes as was the case after the 1990 storm \citep{Acarreta_Sanchez-Lavega_1999, Perez-Hoyos_Sanchez-Lavega_2006} rather than lower them. Thus, it is likely that the wind speed change detected in our measurements indicate a real change in the zonal wind profile, and the thermal wind analysis by \cite{Achterberg_etal_2012DPS} gives confidence to this scenario.

We propose two mechanisms to explain this wind speed change. First, the storm may have acted as a source of anticyclonic vorticity. Figure~15 shows that the primary difference between pre- and post-storm wind profile is the addition of anticyclonic shear around the storm. The growth in area of the storm implies horizontal divergence of the flow, which leads to anticyclonic vorticity when the divergence takes place on a rotating planet. Second, the cumulus storm should have released a substantial amount of latent heat at the cloud-base. Consequently, the storm must have increased (decreased) the latitudinal temperature gradient $dT/dy$ to the south (north) of the storm center. In response to the latitudinal temperature gradient change, the vertical shear in the region should have been modified according to the thermal wind equation \citep{Holton_1992}. If we assume that the wind speed at the cloud condensation level stayed fixed during the course of the storm, the cloud-top wind speed would increase to the north and decreases to the south, consistent with the observed wind speed change. The tropospheric warming detected by \cite{Achterberg_etal_2012DPS} at the storm latitude supports this scenario.

\subsection{Anticyclone Evolution}
Even though the initial size of the AV was comparable to that of Jupiter's Oval BA, the AV had shrunk considerably since its formation. Figure~16 presents latitude-longitude projected CB2-MT2-MT3 composite color maps of the AV to illustrate its size and morphology change over the course of 10~months in 2011. Initially, the AV was outlined by bright clouds that reflected in all three wavelength bands thus appeared white in this color scheme. The center of the vortex appear deep blue, indicating less scattering by deep cloud and increased scattering by high haze. In the subsequent images, the bright outlining clouds gradually fade, while the core scatters more light in both MT2 and CB2, and exhibit a lighter blue color. It is also apparent that the size of the vortex shrank with time. 

In Fig.~17a, we show the deviation in the AV's longitudinal drift from Eq.~3, the mean longitudinal motion. The figure shows the longitudinal extent of the vortex, demonstrating that the deviation from the mean motion rarely exceeds the longitudinal size of the vortex until September~2011. Figure~17b illustrates the change in the longitudinal size of the vortex; it shows that the outer rim of the cloud reached the maximum extent in late April and early May of 2011. Figure~16 panel 2011-04-25 shows the morphology of the vortex when we measured the maximum longitudinal size. Figure~17b also shows that the size of the AV's core fluctuated in longitudinal size between 1.9\degree - 6.6\degree. The images analyzed in Fig.~17 are navigated to better than 0.2\degree~uncertainty in both latitude and longitude; however, the uncertainty in the AV size and center coordinates are dominated by the nebulous appearance of the AV's morphology. A transition of the cloud contrast often spans 1\degree~in latitude and longitude, thus we estimate the uncertainty in our AV coordinate measurements presented in Fig.~17 to be $\pm$1\degree.

As the size of the vortex decreased, the center of the AV has shifted to the north as apparent in Fig.~16. Figure~17c presents the center latitude and the latitudinal extent of the AV's core and outlining region as a function of time. The figure shows both the core and the outlining region of the AV gradually shrank in 2011. It also shows that the AV's center latitude shifted to the north in late August by a distance greater than the size of the core. Between January and July 2011, the center latitude fluctuated between 34.4\degree N and 36.3\degree N; in contrast, between August and October the center latitude was between 36.3\degree N and 37.5\degree N. 

The northward shift of the AV is surprising because an anticyclonic vortex normally drifts equatorward due to the $\beta$-effect. Neptune's Great Dark Spots are examples of anticyclones that drift toward the equator \citep{Hammel_etal_1989NepWind, Sromovsky_1991, Sromovsky_etal_1993}. The poleward drift of the AV suggests that the effective $\beta$ at the latitude is negative, i.e., the background potential vorticity gradient changes sign with latitude. Analyzing the curvature of the zonal wind profile reveals that the wind profile in the region violates the Rayleigh-Kuo (RK) stability criterion, i.e., $\beta -u_{yy}$ does not change sign \citep{Kuo_1949}. Here, $\beta$ is the latitudinal gradient of Coriolis parameter, $u_{yy}$ is the second north-south derivative of zonal wind profile $u(y)$, and $y$ is the north-south distance. Figure~17d compares the observed wind profiles on May 7, 2008 and August 5, 2011 against the maximum curvature allowed by the RK criterion (i.e., $u_{yy} = \beta$). The figure shows that, around 35\degree N, both zonal wind profiles violate the RK criterion, i.e., their curvatures are greater than $\beta$. A qualitative way to reconcile the stable background profile and the stability criterion violation is to invoke that the effective $\beta$ is negative in the region, and the observed poleward drift of the AV is consistent with a negative effective $\beta$. The modification of $\beta$ can occur through the underlying vertical shear structure, which has an effect analogous to the topographic $\beta$ effect \citep{Dowling_1995AnnRev}. Numerical modeling of hte storm's morphological behaviors by \cite{Sanchez-Lavega_etal_2011Nature} reach a similar conclusion. Figures~17e and 17f present the zonal wind shear and the zonal wind curvature. Our results are consistent with similar analyses by \cite{Sanchez-Lavega_etal_2012gws}, who showed that the zonal wind profile at the storm's latitude violated the RK criterion before the storm in 2009; our results demonstrate that the violation persists after the storm. 

As the AV's center latitude shifted to the north, its zonal drift acquired a fast eastward motion as shown in Figs.~3b and 17a. Until August~2011, while the AV's center latitude stayed around 35\degree N, its motion was reasonably well-described by the constant $-$8.4~m$^{~}$s$^{-1}$ (i.e., \emph{westward}) drift rate (Eq.~\ref{e:AV_motion} and dashed line in Fig.~3b). Comparing Fig.~17a and 17c reveals that the small increase in the eastward motion around May-June of 2011 (shown in 16a) corresponds to the slight northward shift in the center latitude. After August 2011, it gradually changed its motion such that it drifted eastward. The drift rate between August~25 and September~6 was $+$2.4~m$^{~}$s$^{-1}$, and by October, the drift rate increased to $+$18~m$^{~}$s$^{-1}$. The mean drift rate between August 25 and October 22 is $+$9.5~m$^{~}$s$^{-1}$ during which time the central latitude stayed around 37\degree N. In comparison, on August~5, 2011, the zonal wind speed at 35\degree N and 37\degree N are $-$13~m$^{~}$s$^{-1}$ and $+$30~m$^{~}$s$^{-1}$, respectively. Figure~17d demonstrates that the increase in AV's eastward drift rate roughly follows the zonal wind profile.

\section{Discussion}
We have reported the evolution of Saturn's latest planet-encircling storm of 2010-2011 using two of the instruments aboard Cassini spacecraft, ISS and RPWS. Our observational coverage enabled us to study the preconditions of the storm, in which we show that the storm erupted out of a previously known feature called the String of Pearls (SoPs). After the storm erupted on December~5, 2010, the outburst grew and engulfed the entire latitude zone. By January 11, 2011, the storm developed a well defined structure with three primary parts as shown in Fig.~5. The western most feature is the particularly bright \emph{head} that propagated at an average rate of $-$2.79$\pm$0.08 \degree~day$^{-1}$ ($-$26.9$\pm$0.8~m~s$^{-1}$), and its longitudinal position is well described by Eq.~\ref{e:head_motion}. A bright body of cloud followed to the east of the head, which is bounded to the east by a new large anticyclonic vortex (AV) that was spawned by the storm. To the east of the AV, a turbulent pattern of clouds formed. The AV maintained a mean drift speed of $-$0.85\degree~day$^{-1}$ ($-$8.4~m$^{~}$s$^{-1}$) between December~24, 2010 and June~14, 2011; during this period, the center of AV did not deviate from Eq.~\ref{e:AV_motion} by more than 10\degree~in longitude (Fig.~17a). We performed cloud-tracking wind measurement on the AV, and showed that its tangential wind speed reached 100~m~s$^{-1}$; we also measured that the relative vorticity within the vortex to be $-$6$\pm$1$\times$10$^{-5}$~s$^{-1}$.

With the east-west diameter of 12,000~km, the new anticyclonic vortex became greater in size than any tropospheric vortex previously seen on Saturn. The rapid growth and shrinkage of the new anticyclone is in a stark contrast to the steadiness of anticyclonic vortices on Jupiter. Oval BA today is a result of many dynamical changes over the last $\sim$70 years; however, it has mostly been growing in size through mergers, and the precursor vortices also stayed steady for over 50~years until the mergers. Compared to Jupiter, Saturn has been known to harbor far fewer vortices \citep{Vasavada_etal_2006}. Our analysis suggests that Saturn's inability to maintain even the largest of the vortices contributes to this difference between Jupiter and Saturn. Even though the new Saturnian anticyclone emerged in a highly disturbed region, we do not expect the turbulence to have contributed to its inability to stay large because Jupiter's GRS, for example, resides in the most turbulent latitude of Jupiter and yet it has persisted at least since 1879, and possibly nearly 400 years \citep{Hooke_1665, Beebe_Orton_West_1989}. Lastly, we note that a stratospheric vortex larger in size than the AV emerged in the aftermath of the storm, which persists as of late 2012 \citep{Fletcher_etal_2012DPS}; its future evolution will be of an interest in the context of Saturnian vortex dynamics.

The storm underwent a major change in its dynamics after around June~20, 2011 when the storm's head collided with the anticyclone after the anticyclone's longitude trailed 360\degree~longitude behind the core. After the collision, the storm's convective activity displayed a major decline. The SED signals emitted by the storm's lightning activities followed the growth and demise of the cumulus storm observed in the ISS images until the head-AV collision in mid-June. The head-AV collision marked a sharp decline in the SED activities detected by the RPWS. In late August 2011, the SED signal resurged for 9~days, and we identified the source clouds in the ISS images. The RPWS detected two more weak SED activities for the remainder of 2011 during September 30 - October 6, and December 26-28.

The storm left the region between 25\degree N and 40\degree N in a highly disturbed state. After the storm, a region that appears particularly dark in CB2 channel emerged between 31\degree N - 38\degree N latitude spanning up to 180\degree~in longitude. After the storm, this region exhibited the lowest albedo on Saturn in the CB2 channel. The region surrounding the dark region exhibited billowing cloud patterns that are similar to mid-latitudes of Jupiter. 

The storm also altered the zonal mean wind profile of the storm latitudes. Our cloud-tracking measurements reveal that the zonal mean cloud motions accelerated by 35~m$^{~}$s$^{-1}$ to the north of the storm around 38\degree N and decelerated by 30~m$^{~}$s$^{-1}$ to the south around 31\degree N. This change in the zonal wind speed is supported by the tropospheric warming detected by \cite{Achterberg_etal_2012DPS}. The warming is consistent with the latent heat released by the storm. Although the change in the cloud motion could be explained by changes in the cloud altitude and vertical shear, we believe that a real wind speed change was caused by the tropospheric warming.

The westward jets at $\pm$35\degree~latitudes have often been the sites of storm activity. Saturn's 35\degree N region harbored several small cumulus storms during the Voyager flybys in 1980-81 \citep{Hunt_etal_1982, Sromovsky_etal_1983}. Earlier in the Cassini mission, similar behaviors, albeit at much smaller scales than the northern-hemisphere planet-encircling storms, were seen in the lightning storms observed in ``Storm Alley,'' at 35\degree S latitude. Like the new storm at 35\degree N, the cumulus events in Storm Alley drifted west and spawned anticyclonic ovals that slowly separated from the source region \citep{Porco_etal_2005, Dyudina_etal_2007}. Unlike in the new storm, the anticyclones drifted west relative to the source, and the source was intermittent. The area of the source region, intensity of the lightning activities, and the size of the anticyclonic vortices were an order of magnitude smaller in the southern storms than in the new northern storm.

The five planet-encircling storms previously seen on Saturn alternated their location between equatorial and mid-latitudes \cite{Sanchez-Lavega_1994}. The last event of 1990 was equatorial \citep{Sanchez-Lavega_etal_1991, Sanchez-Lavega_1994, Beebe_etal_1992, Westphal_etal_1992}. The new storm of 2010-2011 followed the previous pattern of events by erupting in mid-latitude. Comparing the earlier storms observed from Earth with the unprecedented details of the present storm revealed by our present study is not straightforward. However, the large-scale structures seen by the ground-based observations of \cite{Sanchez-Lavega_etal_2012gws} are consistent with the results presented here, and give confidence in our ability to compare duration and the spatial structures of the present storm with the past ones. If we take December 5, 2010 as the beginning and June~20, 2011 as the end of the present storm, it lasted for 201~days, longer than any of the five previously recorded storms \citep{Sanchez-Lavega_1994}. Interestingly, the previous longest storm, which lasted for 150 days in 1903, erupted at 30$\pm$2\degree N, a latitude similar to the present storm. On Earth, one of the factors that contribute to the longevity of a thunderstorm is the lack of vertical shear \citep{Emanuel_1994}. The two long-lasting cumulus storms may indicate the lack of vertical shear at their latitudes.

Episodic convective outbursts such as the Saturnian planet-circling storms suggest the presence of a mechanism that allows a large build-up of convective available potential energy (CAPE) between the large outbursts. At the depth of 10-bar at the predicted water condensation level, the static stability is believed to be enhanced by the latent heat release \citep{Weidenschilling_Lewis_1973, Sugiyama_etal_2006}. This layer of enhanced static stability could act as a lid for any convective activities that initiate underneath and store CAPE. An episodic cumulus storm can erupt when the internal heat flux deposits sufficient CAPE to break through the water condensation layer; this instability could be triggered spontaneously or modulated by seasonal effects. Perhaps the String of Pearls forms when Rossby waves are excited in the water condensation layer. \cite{Sayanagi_Showman_2007} demonstrated that Rossby waves excited at the 10-bar level can affect the cloud-top level. This possibility can be tested if simultaneous observations by VIMS and ISS reveal vertical offset in the SoPs phase; as those instruments sense different altitudes, comparing their images should reveal the vertical structure of the SoPs. If SoPs is indeed a Rossby wave, the ISS images should show SoPs phase shifted to the west compared to that in the VIMS images. This will be a topic for future investigation.

Based on the records of past events, we expect the disturbance to have lasting effects on Saturn's northern hemisphere. As a point of comparison, the 1990 storm left the northern hemisphere of Saturn disturbed for the rest of the decade, and many activities were recorded until 1997 \citep{Sanchez-Lavega_etal_1993a, Sanchez-Lavega_etal_1996, Sanchez-Lavega_etal_1999}. The continuing disturbances of the 2010-2011 storm have been confirmed through ground-based telescopic observations and Cassini RPWS measurements. For example, in early April 2012, the RPWS detected electrostatic discharge signals emitted by a new storm. Cassini CIRS and ground-based infrared observations also continue to show lasting activities in the stratosphere \citep{Hesman_etal_2012DPS, Fletcher_etal_2012DPS, Hesman_etal_2012, Fletcher_etal_2012}, and show a large stratospheric hot spot that has been termed the~``beacon.'' Like the 1990 storm, the aftermath of the latest storm should have an effect that may last for up to a decade, and a continuing monitoring of Saturn from the orbiting vantage point of Cassini spacecraft should reveal further details of the dynamic event.

\textbf{Acknowledgement:}
Our work was supported by the Cassini-Huygens mission, a cooperative project of NASA, ESA, ASI, managed by JPL, a division of the California Institute of Technology, under a contract with NASA. The authors thank the two anonymous reviewers for their very constructive comments.
\label{lastpage}


\def\ref@jnl#1{{\jnl@style#1}}%
\newcommand\aj{\ref@jnl{AJ}}%
\newcommand\araa{\ref@jnl{ARA\&A}}%
\newcommand\apj{\ref@jnl{ApJ}}%
\newcommand\apjl{\ref@jnl{ApJ}}%
\newcommand\apjs{\ref@jnl{ApJS}}%
\newcommand\ao{\ref@jnl{Appl.~Opt.}}%
\newcommand\apss{\ref@jnl{Ap\&SS}}%
\newcommand\aap{\ref@jnl{A\&A}}%
\newcommand\aapr{\ref@jnl{A\&A~Rev.}}%
\newcommand\aaps{\ref@jnl{A\&AS}}%
\newcommand\azh{\ref@jnl{AZh}}%
\newcommand\baas{\ref@jnl{BAAS}}%
\newcommand\jrasc{\ref@jnl{JRASC}}%
\newcommand\memras{\ref@jnl{MmRAS}}%
\newcommand\mnras{\ref@jnl{MNRAS}}%
\newcommand\pra{\ref@jnl{Phys.~Rev.~A}}%
\newcommand\prb{\ref@jnl{Phys.~Rev.~B}}%
\newcommand\prc{\ref@jnl{Phys.~Rev.~C}}%
\newcommand\prd{\ref@jnl{Phys.~Rev.~D}}%
\newcommand\pre{\ref@jnl{Phys.~Rev.~E}}%
\newcommand\prl{\ref@jnl{Phys.~Rev.~Lett.}}%
\newcommand\pasp{\ref@jnl{PASP}}%
\newcommand\pasj{\ref@jnl{PASJ}}%
\newcommand\qjras{\ref@jnl{QJRAS}}%
\newcommand\skytel{\ref@jnl{S\&T}}%
\newcommand\solphys{\ref@jnl{Sol.~Phys.}}%
\newcommand\sovast{\ref@jnl{Soviet~Ast.}}%
\newcommand\ssr{\ref@jnl{Space~Sci.~Rev.}}%
\newcommand\zap{\ref@jnl{ZAp}}%
\newcommand\nat{\ref@jnl{Nature}}%
\newcommand\iaucirc{\ref@jnl{IAU~Circ.}}%
\newcommand\aplett{\ref@jnl{Astrophys.~Lett.}}%
\newcommand\apspr{\ref@jnl{Astrophys.~Space~Phys.~Res.}}%
\newcommand\bain{\ref@jnl{Bull.~Astron.~Inst.~Netherlands}}%
\newcommand\fcp{\ref@jnl{Fund.~Cosmic~Phys.}}%
\newcommand\gca{\ref@jnl{Geochim.~Cosmochim.~Acta}}%
\newcommand\grl{\ref@jnl{Geophys.~Res.~Lett.}}%
\newcommand\jcp{\ref@jnl{J.~Chem.~Phys.}}%
\newcommand\jgr{\ref@jnl{J.~Geophys.~Res.}}%
\newcommand\jqsrt{\ref@jnl{J.~Quant.~Spec.~Radiat.~Transf.}}%
\newcommand\memsai{\ref@jnl{Mem.~Soc.~Astron.~Italiana}}%
\newcommand\nphysa{\ref@jnl{Nucl.~Phys.~A}}%
\newcommand\physrep{\ref@jnl{Phys.~Rep.}}%
\newcommand\physscr{\ref@jnl{Phys.~Scr}}%
\newcommand\planss{\ref@jnl{Planet.~Space~Sci.}}%
\newcommand\procspie{\ref@jnl{Proc.~SPIE}}%
\let\astap=\aap
\let\apjlett=\apjl
\let\apjsupp=\apjs
\let\applopt=\ao

\bibliography{bibliography}
\bibliographystyle{elsart-harv}

\begin{noindent}
\textbf{Tables and Figures for Sayanagi et al. ``Dynamics of Saturn's Great Storm Throughout its Lifetime 2010-2011''}

\begin{center}
\textbf{Table 1.} Longitude vs. Time of the Head of Storm\\
\begin{tabular}{c c c}
  \hline
  Year/Month/Fractional Day & East Longitude [degree] & West Longitude [degree] \\ 
  \hline
        2010/12/05.9 & 114.1 & 245.9 \\
        2010/12/24.6 & 62.5 & 297.5 \\
        2011/01/02.5 & 37.1 & 322.9 \\
        2011/01/11.8 & 7.30 & 352.7 \\
        2011/02/04.4 & 302.9 & 57.1 \\
        2011/02/25.8 & 242.8 & 117.2 \\
        2011/03/08.0 & 212.8 & 147.2 \\
        2011/04/22.5 & 94.4 & 265.6 \\
        2011/05/05.2 & 55.4 & 304.6 \\
        2011/06/14.2 & 298.6 & 61.4 \\ 
  \hline
\end{tabular}
\end{center}

\bigskip
\begin{center}
\textbf{Table 2.} Longitude vs. Time of the Anticyclonic Vortex (AV)\\
\begin{tabular}{c c c}
  \hline
  Year/Month/Fractional Day & East Longitude [degree] & West Longitude [degree]\\ 
  \hline
        2010/12/24.6 & 75.2 & 284.8 \\ 
        2011/01/02.5 & 62.9 & 297.1 \\ 
        2011/01/11.8 & 54.5 & 305.5 \\ 
        2011/02/23.9 & 6.2 & 353.8 \\ 
        2011/03/08.0 & 356.5 & 3.5\\ 
        2011/03/17.7 & 347.0 & 13.0\\ 
        2011/04/25.8 & 314.0 & 46.0 \\ 
        2011/05/05.2 & 318.7 & 41.3 \\ 
        2011/05/16.7 & 307.5 & 52.5 \\ 
        2011/06/04.4 & 295.9 & 64.1 \\ 
        2011/06/14.2 & 282.2 & 77.8 \\ 
        2011/07/09.5 & 252.5 & 107.5 \\
        2011/07/12.1 & 246.5 & 113.5 \\
        2011/08/02.3 & 224.2 & 135.8 \\
        2011/08/06.4 & 224.7 & 135.3 \\
        2011/08/08.5 & 224.5 & 135.5 \\
        2011/08/09.6 & 222.6 & 137.4 \\
        2011/08/11.1 & 219.7 & 140.3 \\
        2011/08/12.2 & 219.8 & 140.2 \\
        2011/08/13.2 & 218.0 & 142.0 \\
        2011/08/24.2 & 217.4 & 142.6 \\
        2011/09/06.6 & 226.5 & 133.5 \\
        2011/09/16.3 & 227.8 & 132.2 \\
        2011/09/19.9 & 230.8 & 129.2 \\
        2011/10/06.7 & 247.5 & 112.5 \\
        2011/10/09.8 & 252.3 & 107.7 \\
        2011/10/22.6 & 276.6 & 83.4 \\
  \hline
\end{tabular}
\end{center}

\bigskip
\begin{center}
\textbf{Table 3.} Longitudes of the Dark Ovals (DOs)\\
\begin{tabular}{c c c}
  \hline
  Year/Month/Fractional Day & East Longitude [degree] & West Longitude [degree] \\
  \hline
    2011/01/12.5 & 127, 158 & 233, 202\\
    2011/01/22.3 & 215 & 145 \\
    2011/02/06.9 & 306 & 54 \\
    2011/02/25.7 & 32 & 328 \\
    2011/02/26.3 & 231, 245 & 129, 115 \\
    2011/03/09.6 & 231, 238, 243, 252, 257, 300 & 129, 122, 117, 108, 103, 60\\
    2011/03/22.1 & 340 & 20 \\
  \hline
\end{tabular}
\end{center}
\pagebreak
\begin{center}
\includegraphics[width=8.75in, angle=90]{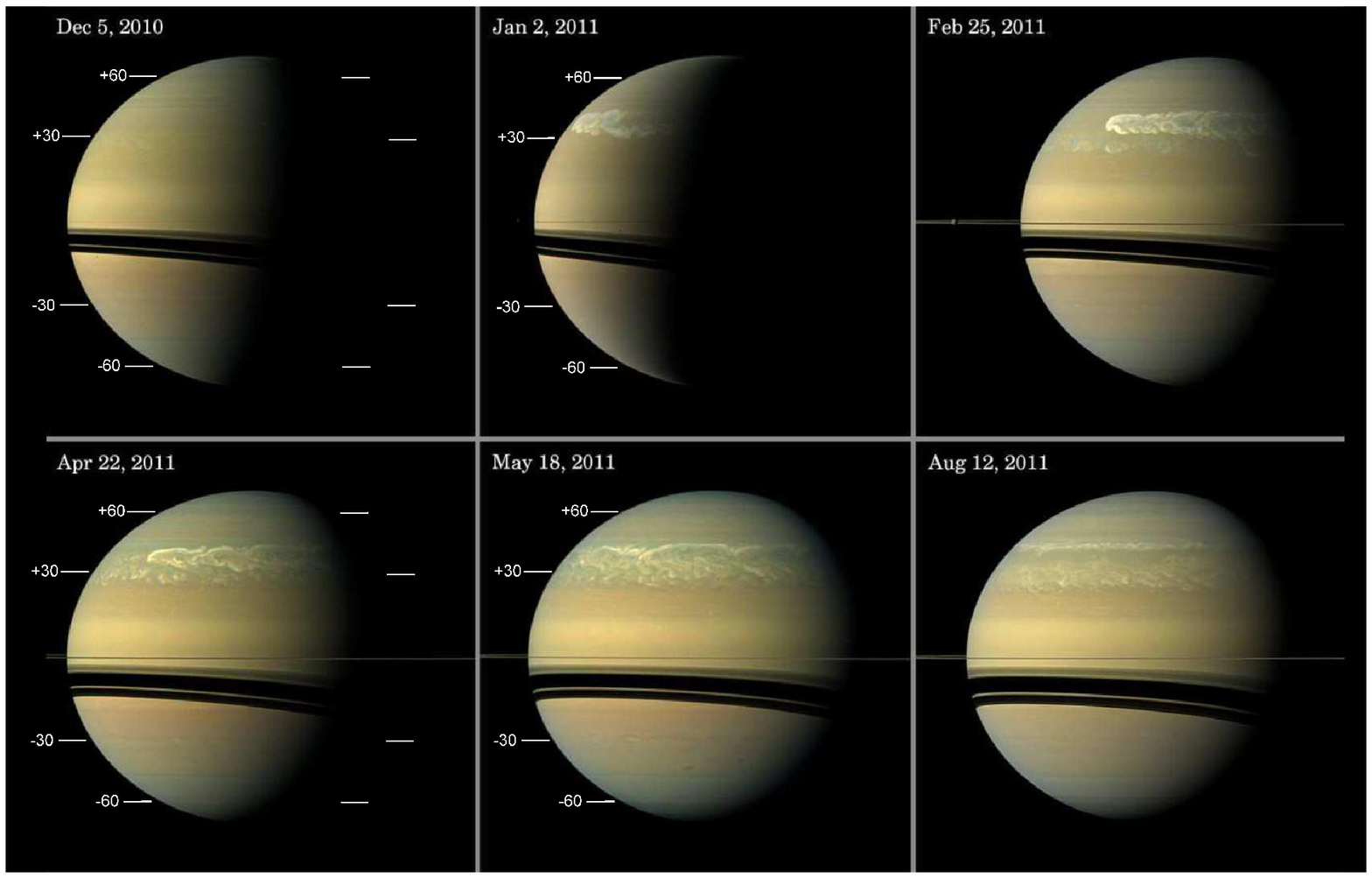}
\end{center}

\noindent \textbf{Figure 1:}
Changing appearance of Saturn in real color. The storm started on December~5, 2010, and the disturbance subsequently engulfed the entire 33\degree N~latitude zone. The planetocentric latitudes are marked on each panel.
\pagebreak

\pagebreak
\begin{center}
\includegraphics[angle=90, width=2.5in]{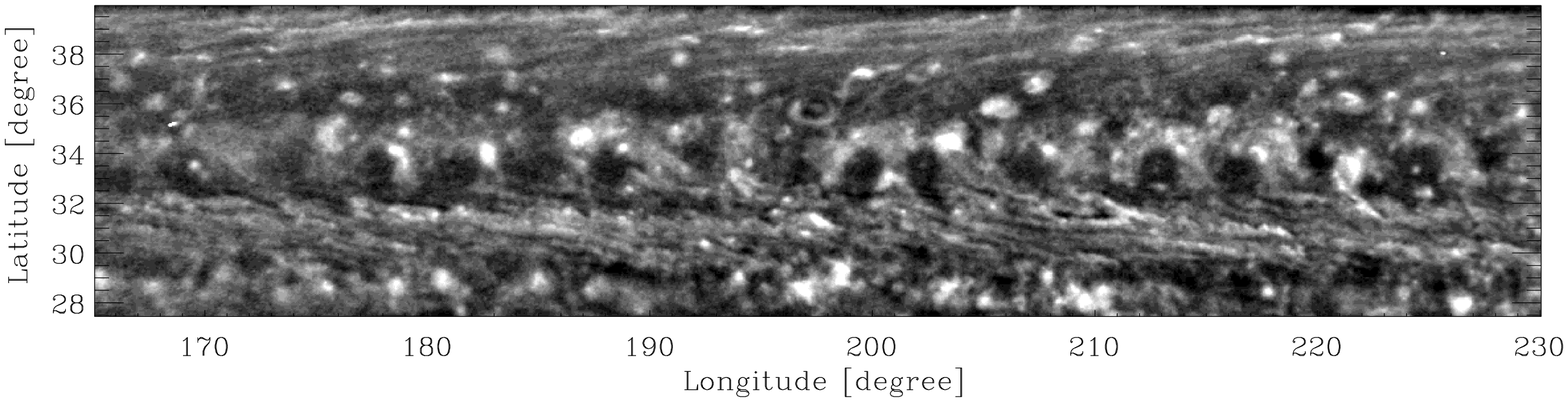}
\end{center}

\noindent \textbf{Figure 2: } 
Latitude-Longitude projected map on March~29, 2008 using Cassini ISS camera's CB2 filter. The 17 dark spots that are aligned at 33\degree N latitude are the SoPs.

\pagebreak
\begin{center}
\includegraphics[width=6.5in]{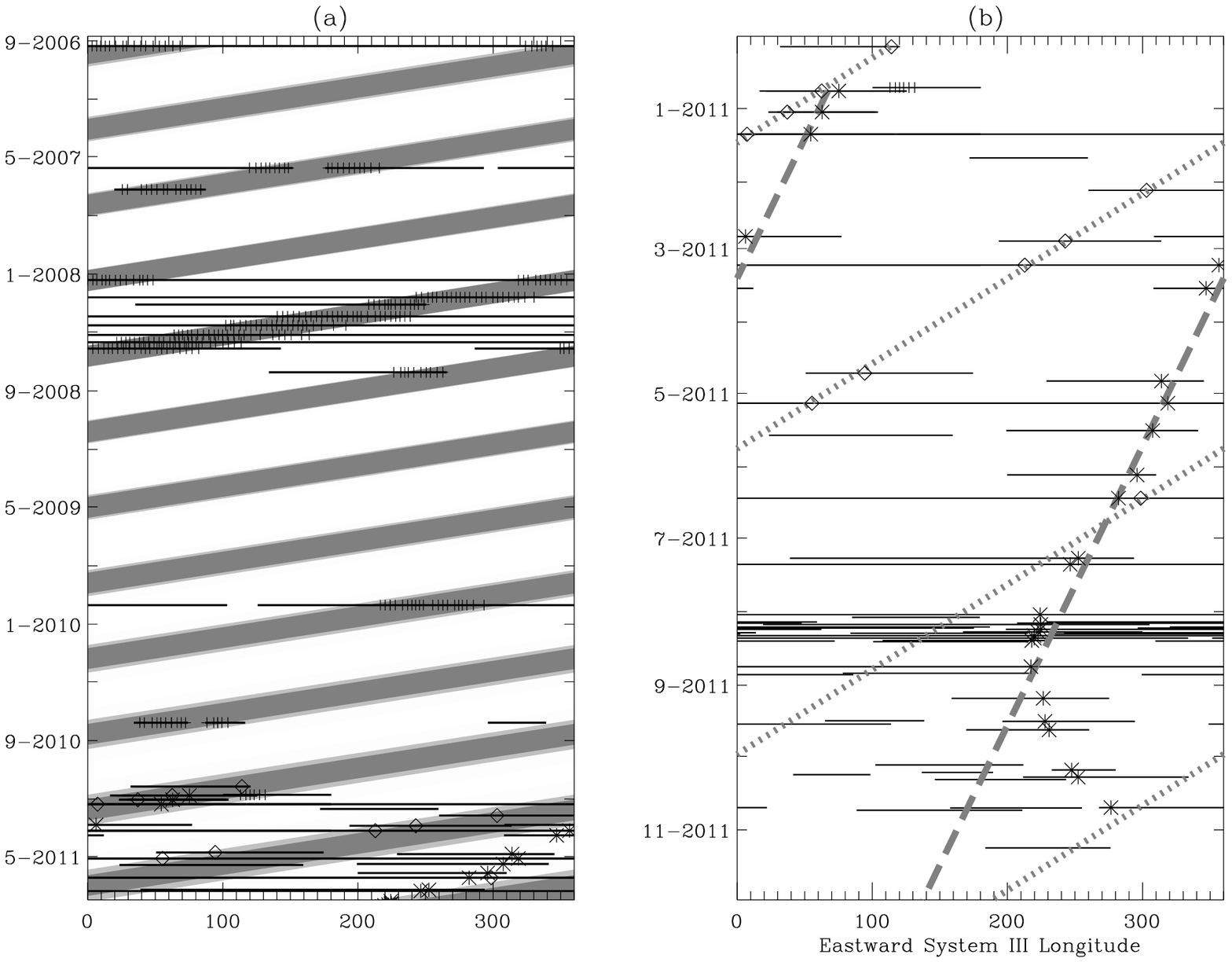}
\end{center}

\noindent \textbf{Figure 3: } 
Motions of the SoPs, storm's head and the AV. The horizontal lines denote the observational coverage at a given time. Pluses (+) denote the longitudes of individual \emph{pearls}. Diamonds ($\diamondsuit$) and asterisks ($\ast$) denote the longitudes of head and AV. In panel (a), the dark shade represents the best-fit motion of the SoPs ($-$22.42~m~s$^{-1}$), and the accompanying half-shade denotes its 1-percent uncertainty envelope. The September~2006 data is VIMS data as documented by Choi et al. (2009). In panel (b), dotted and dashed lines denote the best-fits to the motions of head ($-$26.9~m~s$^{-1}$) and AV ($-$8.4~m~s$^{-1}$), respectively.

\begin{center}
\includegraphics[width=6.0in]{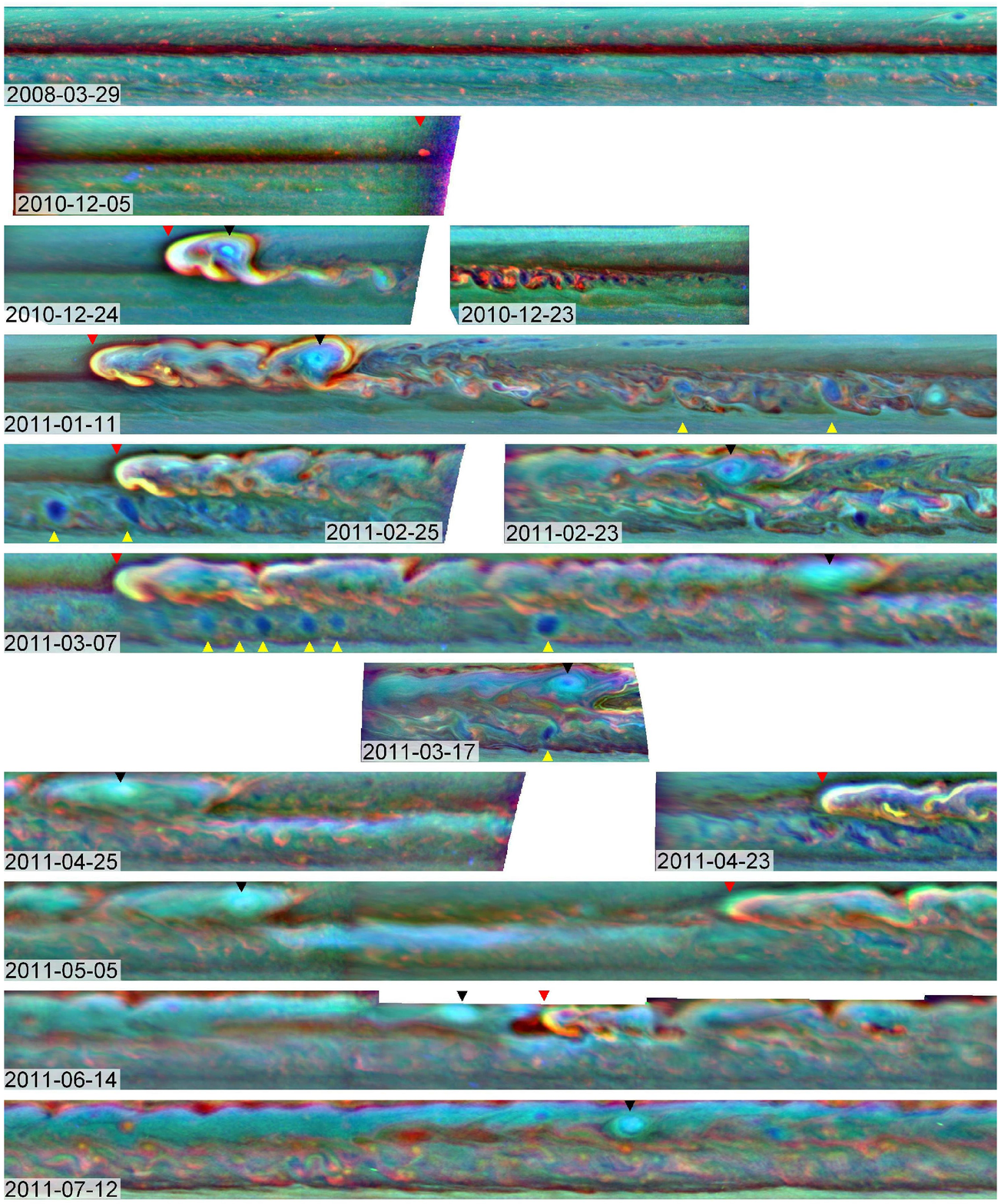}
\end{center}

\noindent \textbf{Figure 4: } 
Cloud morphology before, during and after the storm. The panel IDs denote the image acquisition dates. Each panel is a latitude-longitude projected mosaic and shows 20.5\degree - 38.9\degree N planetocentric latitude. The width of the entire figure represents 200\degree~of longitude. Mosaics are cropped to highlight the spacing between the head (marked with red triangle) and the AV (black triangle). Yellow triangles denote the longitudes of the DOs. Red, green and blue color channels in these composite mosaics are assigned to Cassini ISS camera's CB2 (750~nm), MT2 (727~nm), and MT3 (889~nm) filters, respectively.

\pagebreak
\begin{center}
\includegraphics[width=7.75in, angle=90] {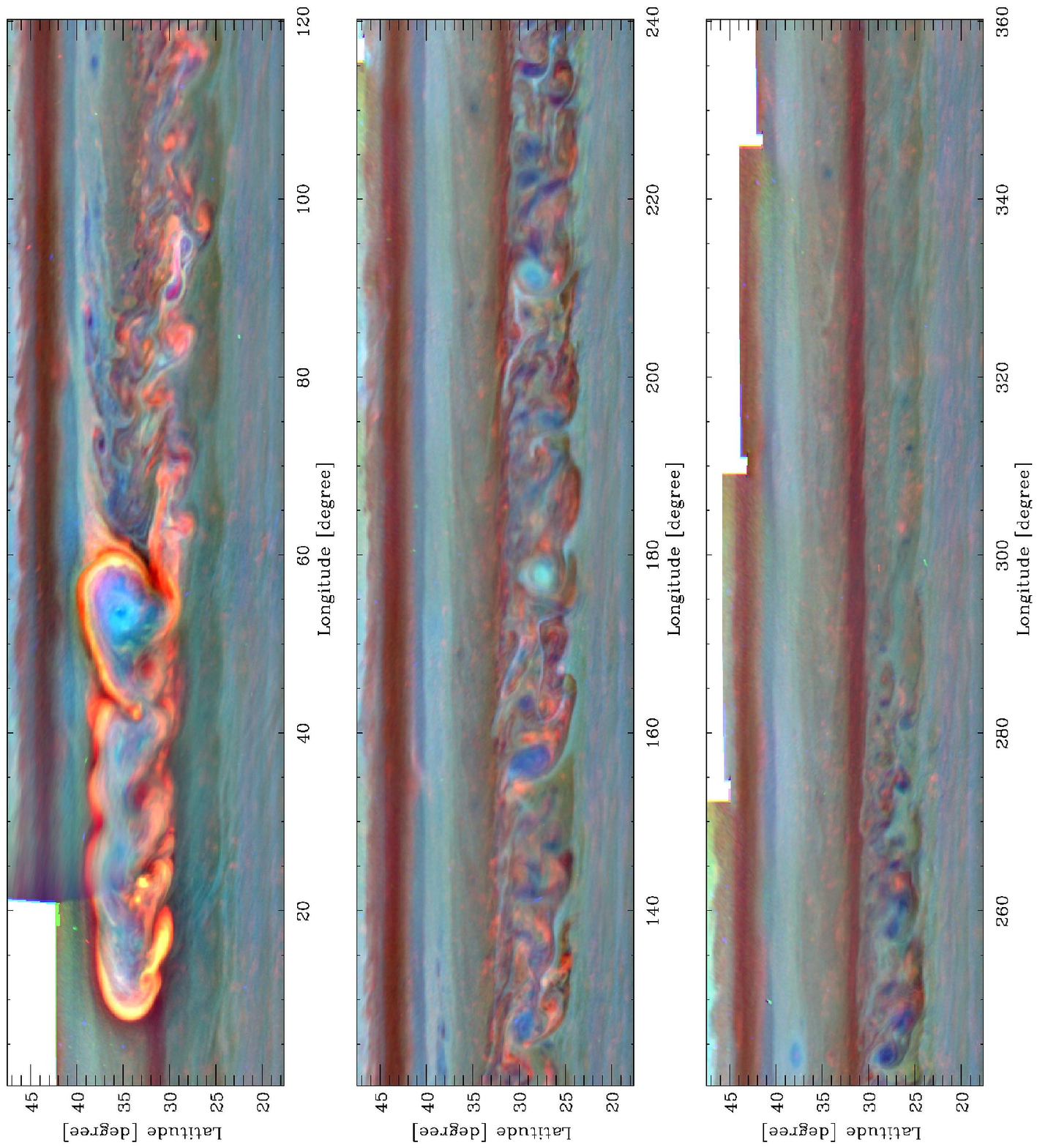}
\end{center}

\noindent \textbf{Figure 5:}
Full-longitude mosaic of Saturn's northern mid-latitudes on January~11, 2011. The red, green and blue color channels are assigned to images acquired using CB2, MT2 and MT3 filters, respectively -- the color scheme is described in the main text.

\pagebreak
\begin{center}
\includegraphics[width=6.25in]{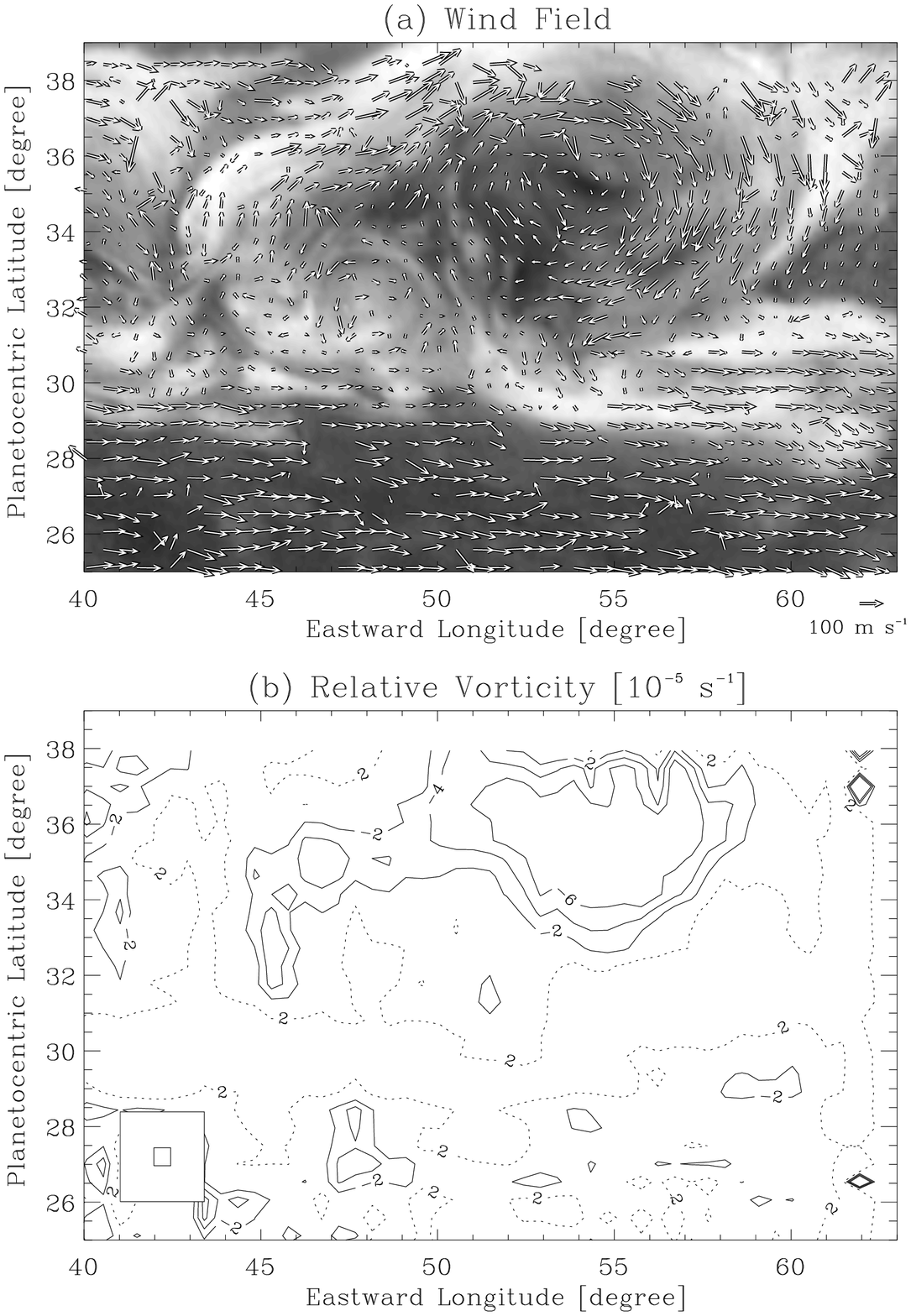}
\end{center}

\noindent \textbf{Figure 6: } 
(a) Wind vectors measured by tracking features within the large anticyclonic vortex that emerged from the storm. The wind was measured on 11~January, 2011. The latitude-longitude projected CB2 image of the vortex is shown in background. (b) The relative vorticity map calculated using the wind field presented in Panel (a). The large square at the lower left of the panel denotes the smoothing scale. The small square denotes the vector grid spacing.

\pagebreak
\begin{center}
\includegraphics[width=3.in]{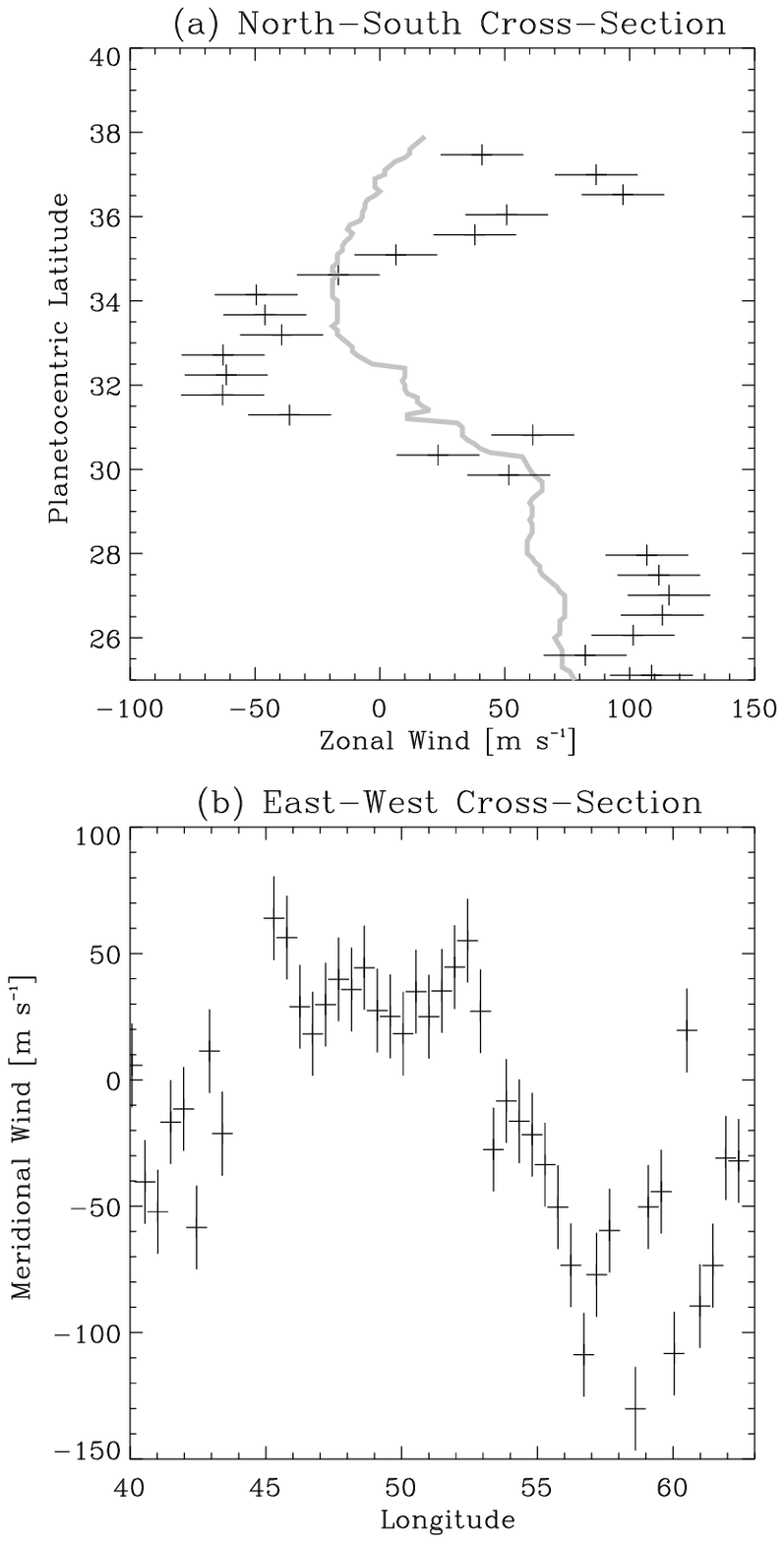}
\end{center}

\noindent \textbf{Figure 7: } 
Cross sections of the anticyclone's wind components on January~11, 2011 using the same data shown in Figure~6. The width of the error bars in both panels, $\pm$16.5~m~s$^{-1}$, was determined using the images' navigational uncertainty. (a): North-south cross section of eastward wind component along the vortex's central longitude (54\degree). The zonal mean (excluding the vortex region) wind of the same data is shown in gray. (b): East-west cross section of northward wind component along the vortex's central latitude (35.5\degree N). 

\pagebreak
\begin{center}
\includegraphics[width=6.in]{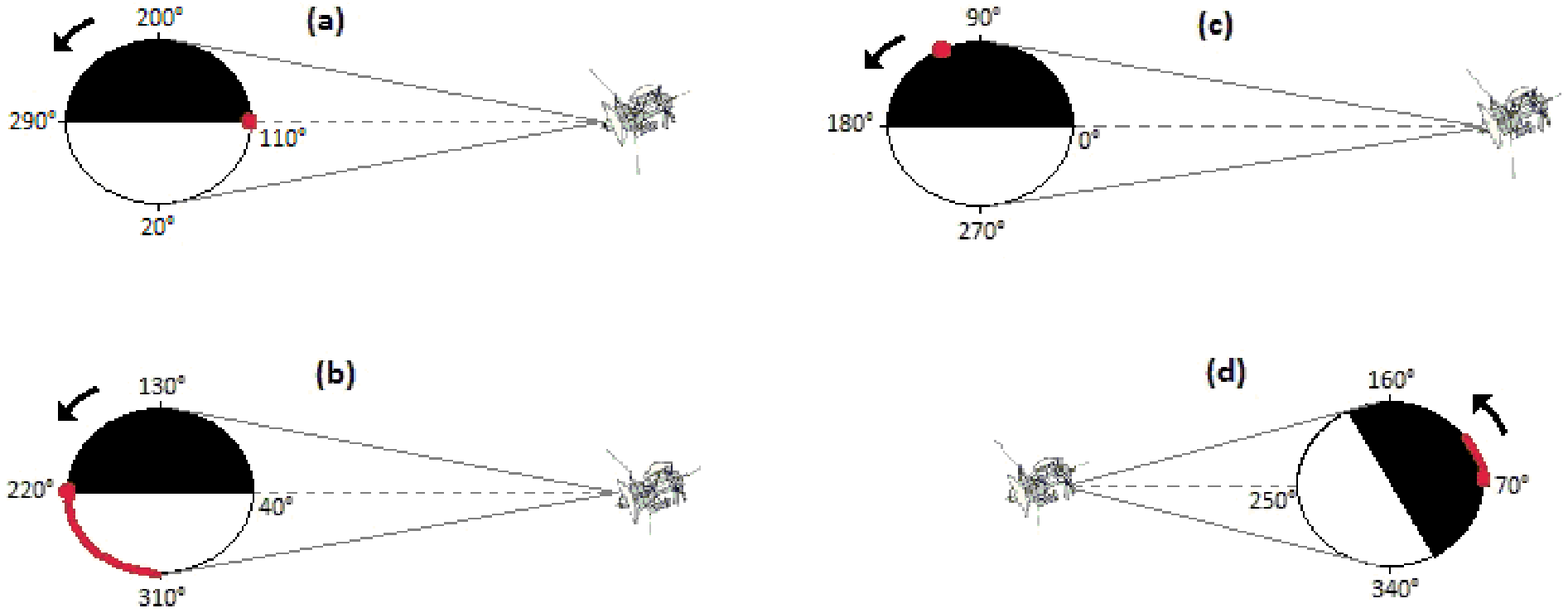}
\end{center}

\noindent \textbf{Figure 8: } 
Each panel illustrates Saturn viewed from above north pole, the Cassini spacecraft, and the SED source. Saturn's dayside and nightside are marked with light and dark shades, and the SED source region is colored red. Black arrow indicates the direction of Saturn's rotation. The visible line-of-sight horizons are marked with the thin solid lines, and the dashed line indicates the sub-spacecraft longitude. Panels (a)-(c) are representative of the most common observation geometry during the period covered in our report, when Cassini spacecraft's orbit had its apoapsis approximately over the evening terminator. Panel (a) shows the geometry on December 7, 2010; a small SED source at the sub-spacecraft longitude of 110\degree. Panel (b) shows the observation geometry of early March 2011 as the SED source, extended over 90\degree in longitude, was starting to rise over the visible morning horizon. Panel (c) depicts the geometry when the SED source sets over the night-side horizon. Panel (d) depicts the geometry after Cassini passed periapsis on December~21, 2010, in which the SED signals reach the spacecraft via the over-horizon effect on the morning side.

\pagebreak
\begin{center}
\includegraphics[width=7.5in]{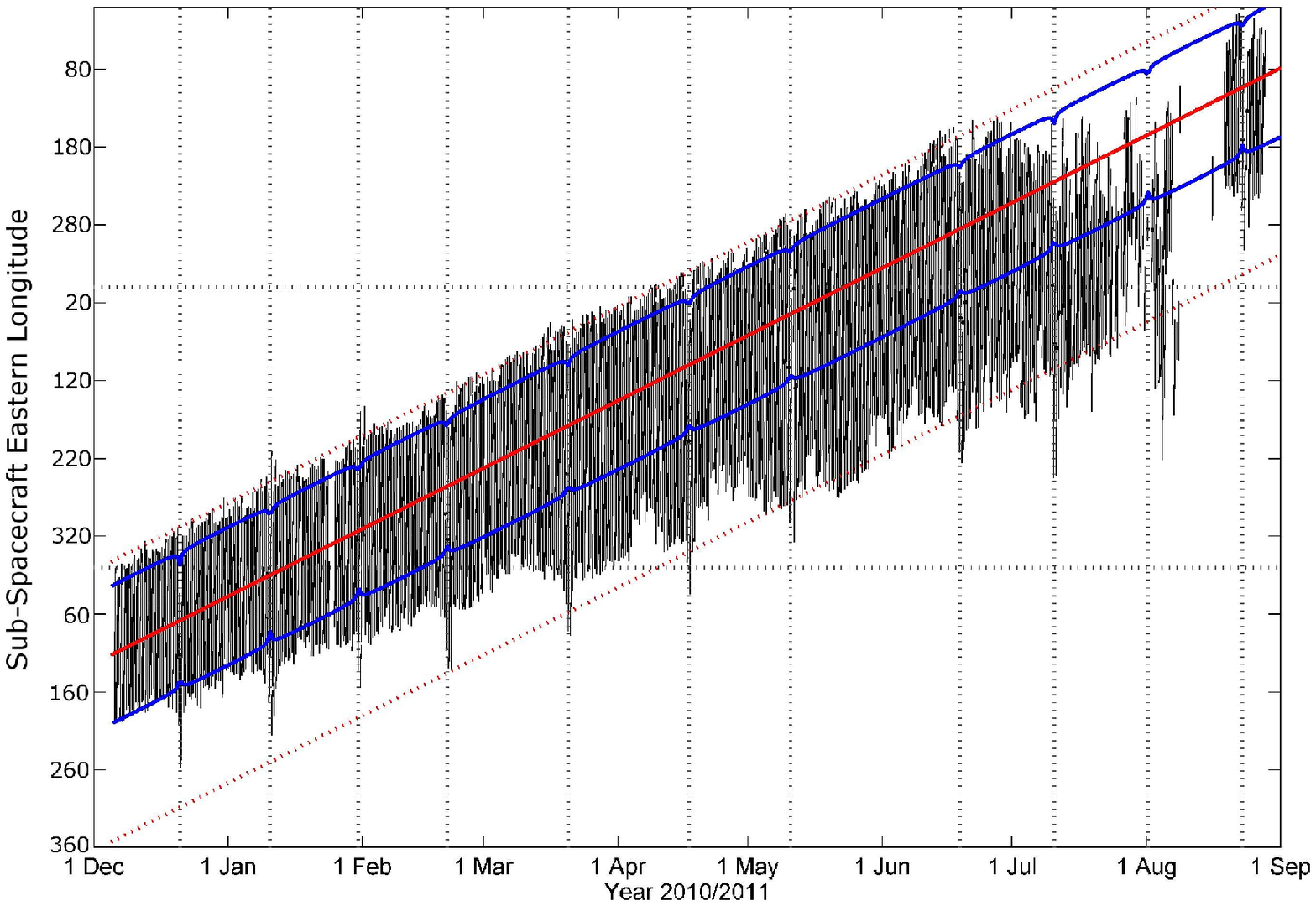}\\
\end{center}

\noindent \textbf{Figure 9: } 
Range of subspacecraft longitudes while SED signals are detected as a function of time. During the course of the measurement, the storm's head circumdrifted the latitude circle three times, thus the longitude axis (i.e., vertical axis) wraps around the latitude circle three times. Each vertical black line denotes the range of sub-spacecraft longitude during which the RPWS instrument detected the SED signal. The inclined red line shows the position of the storm head as given by Eq.~2. The storm's head is in the spacecraft-facing hemisphere while the sub-spacecraft longitude is between the two blue lines. The upper dotted red line marks 120\degree to the west of the storm's head, approximate point where the SED source sets below the radio horizon. The lower dotted red line denotes 240\degree~to the east of the storm head, i.e., a single Saturn rotation is denoted by the width between the dotted red lines. The 11 dotted vertical gray lines indicate Cassini's periapsis passes.

\pagebreak
\begin{center}
\includegraphics[width=7.5in]{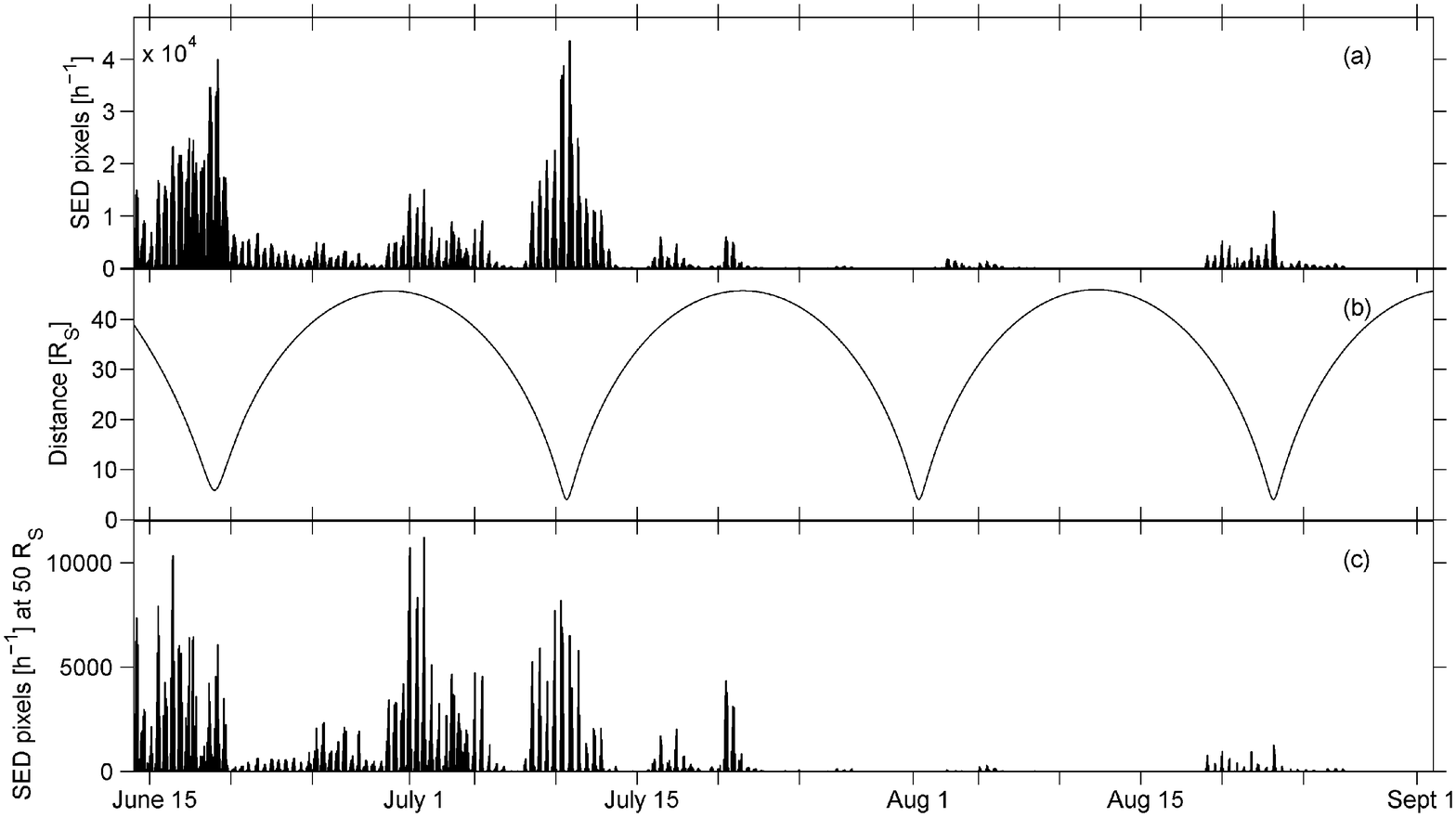}\\
\end{center}

\noindent \textbf{Figure 10: } 
Panel (a): The number of SED pulse detected as a function of time. Panel (b): Spacecraft distance from Saturn -- the vertical axis is in Saturn Radius $R_S$. Panel (c): The number of SED detections scaled with the distance from Saturn -- the SED detections are normalized to 50~$R_S$ distance.

\pagebreak
\begin{center}
\includegraphics[width=7.75in, angle=90]{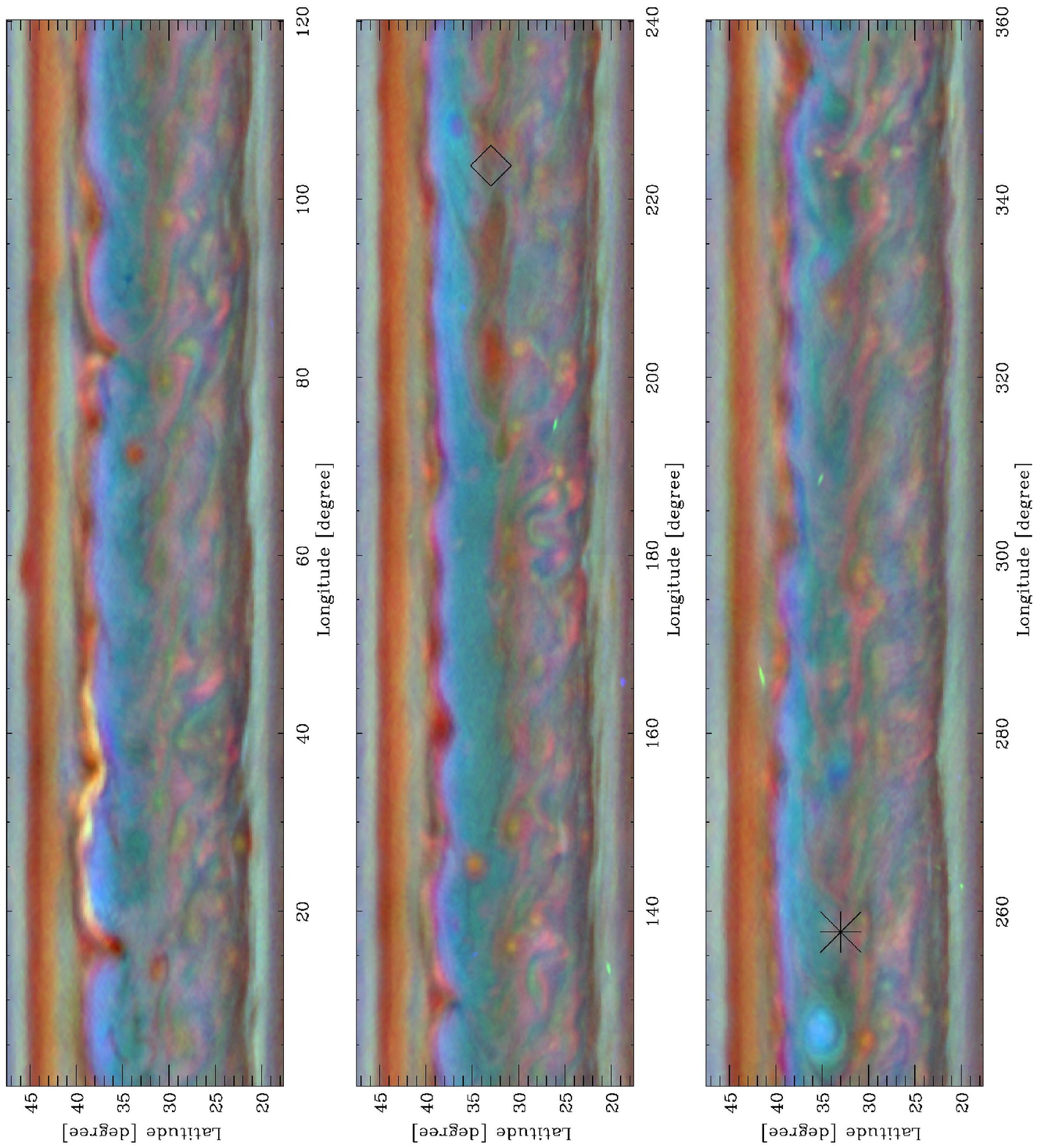}
\end{center}
\noindent \textbf{Figure 11: } 
Full-longitude CB2-MT2-MT3 mosaic of Saturn's northern mid-latitudes on July~12, 2011, after the head-AV collision. The color scheme is the same as in Fig.~5. The longitudes of the storm's head (as predicted by Eq.~(2), even though it is no longer present) and the AV (Eq.~(3)) are marked by a diamond and an asterisk, respectively. The bright cloud centered at around 30\degree~longitude is near the longitudes of SED activities expected from the RPWS measurements shown in Fig.~9.

\pagebreak
\begin{center}
\includegraphics[width=7.75in, angle=90]{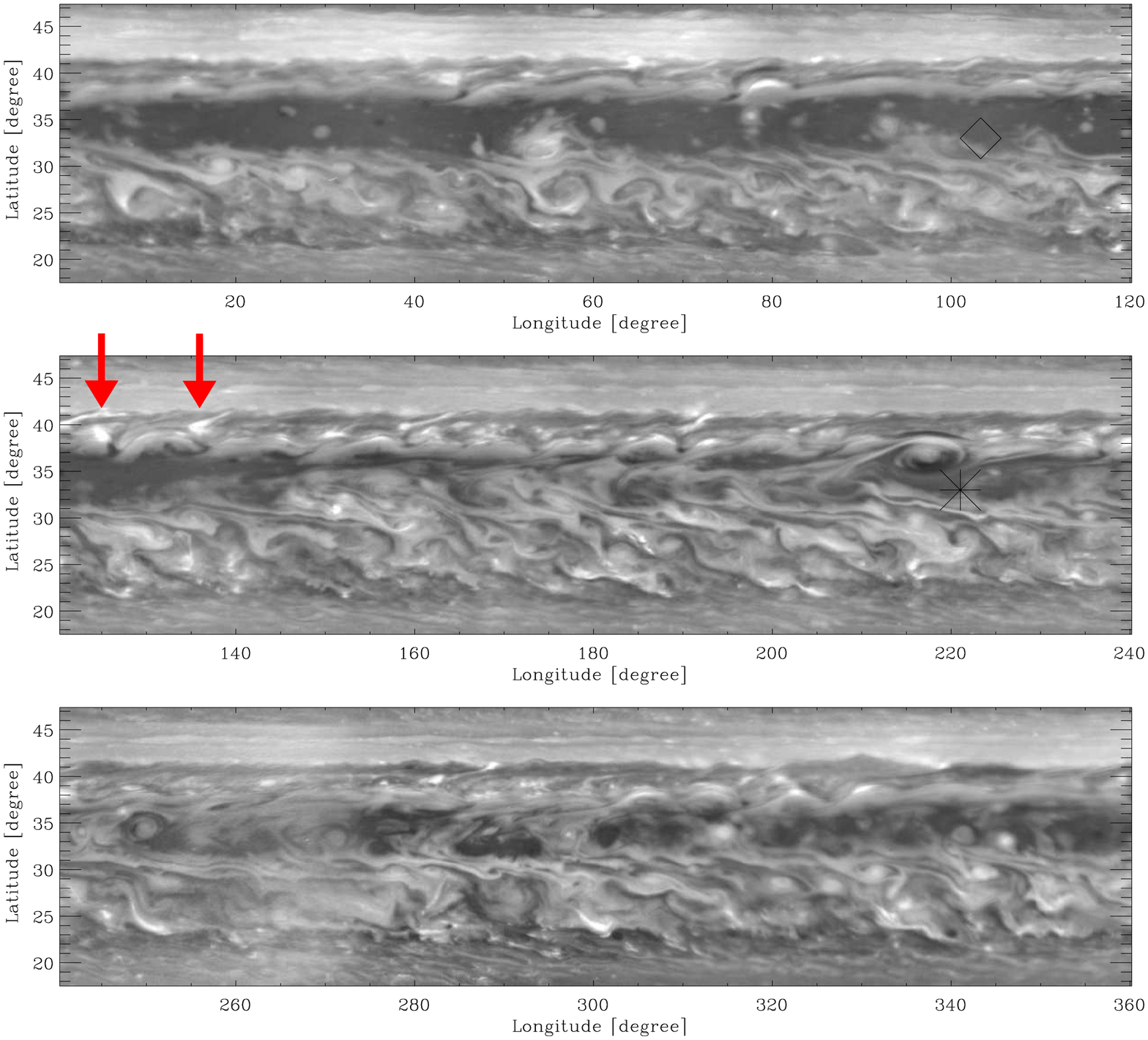}
\end{center}
\noindent \textbf{Figure 12:} 
Full-longitude CB2 mosaic of Saturn's northern mid-latitudes on August~24, 2011, during the resurgence in the SED activities. The predicted longitudes of the storm's head (which is no longer present) and the AV are marked by a diamond and an asterisk, respectively. The bright clouds marked by the arrows are near the longitudes of SED activities expected from the RPWS measurements shown in Fig.~9.

\pagebreak
\begin{center}
\includegraphics[angle=90, width=3.5in]{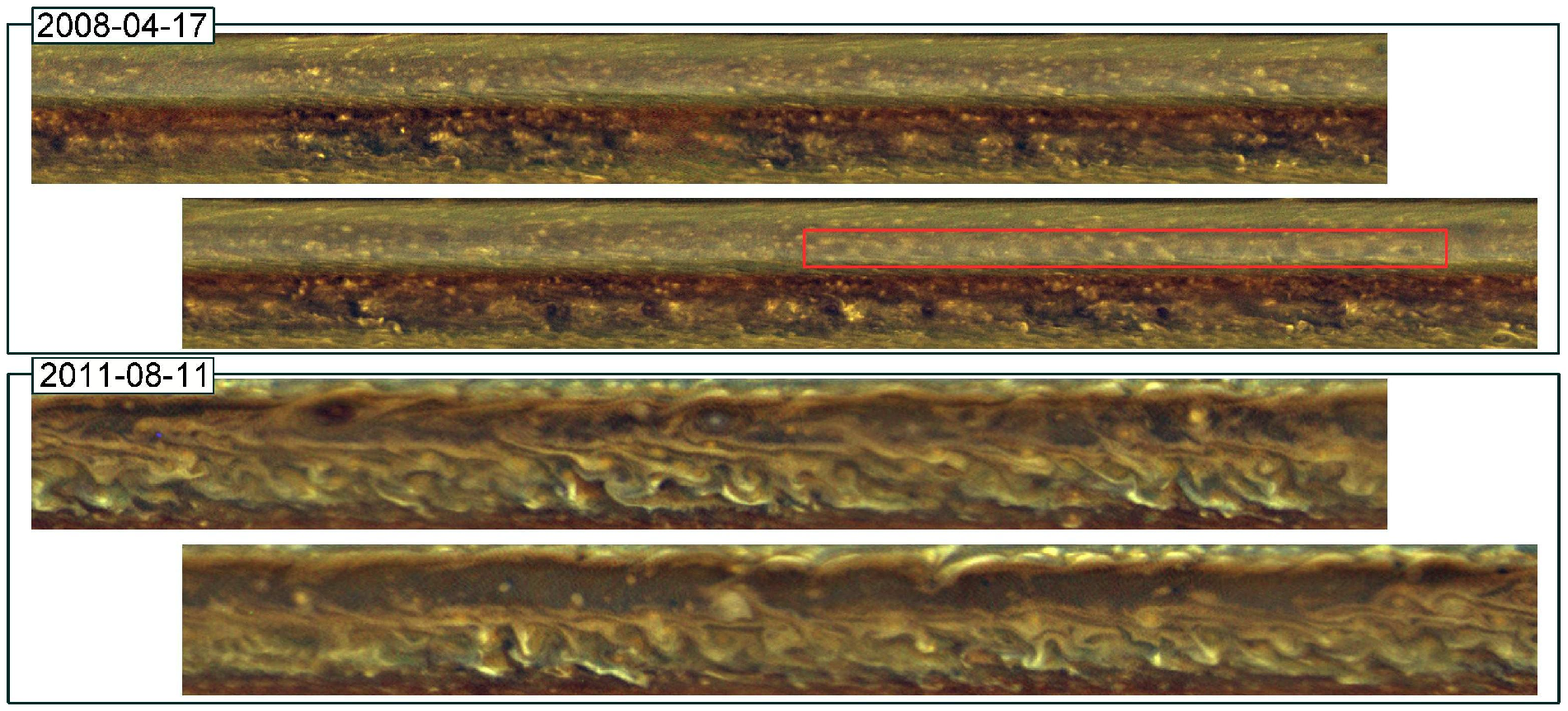}
\end{center}

\noindent \textbf{Figure 13: } 
Comparison of cloud morphology before and after the storm. Each panel shows a latitude-longitude projected, enhanced color mosaics of Saturn between 25\degree - 45\degree N planetocentric latitude. The contrast was separately adjusted in red-, green- and blue-filtered images to construct this composite image. The top two panel present the full-longitude mosaic before the storm on 17~April 2008: the first panel shows $-$180\degree - 0\degree~System III east longitude, and the second shows 0\degree - 180\degree, in which the SoPs is visible inside the red box. The bottom two panels present the same view after the storm on 11~August, 2011. The dark region discussed in the text is prominent in the bottom panel.

\pagebreak
\begin{center}
\includegraphics[width=6.in]{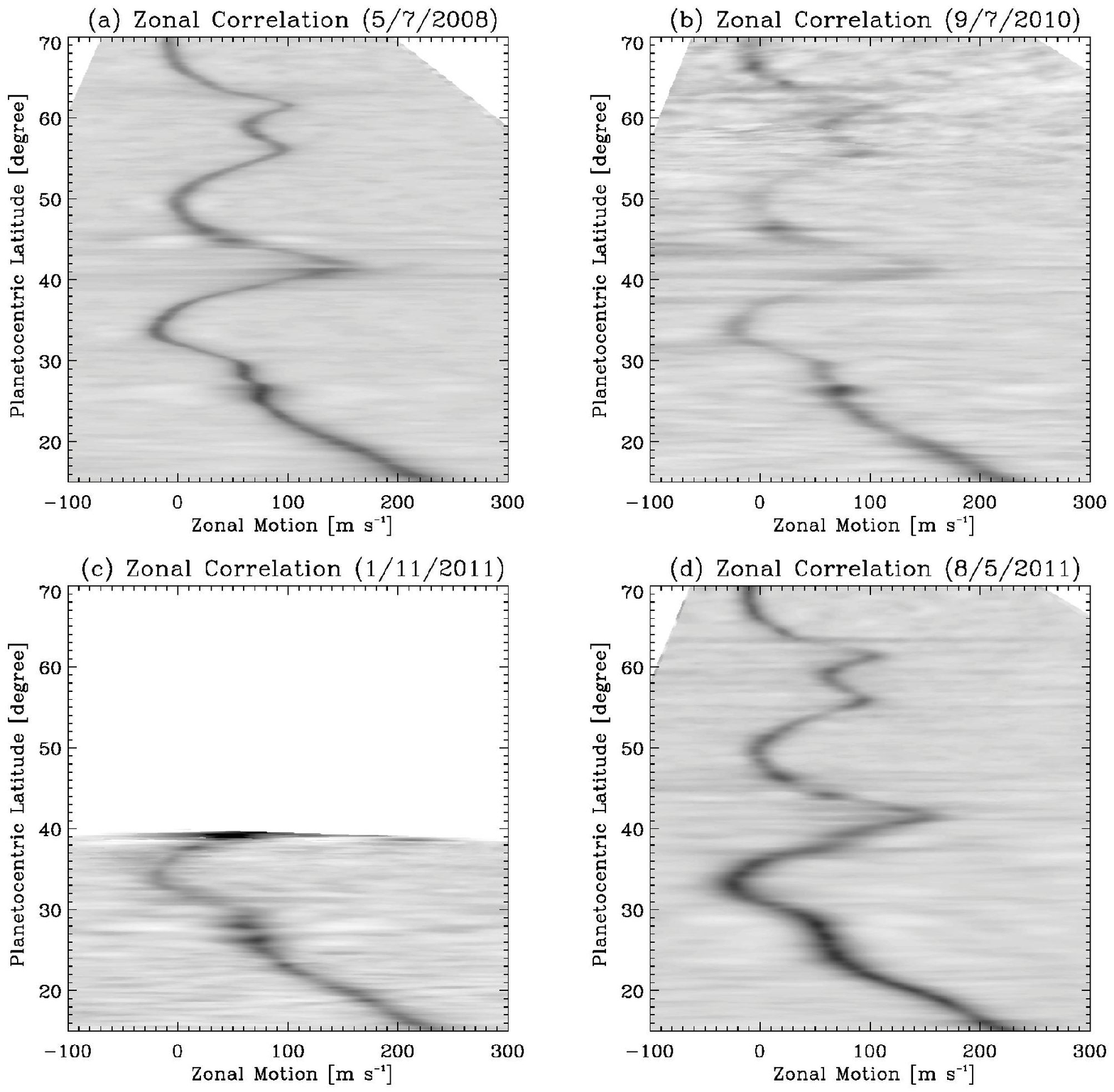}
\end{center}

\noindent \textbf{Figure 14: } 
Zonal wind measurements before, during and after the storm's convective activities. All images analyzed here are captured using the CB2 filter. Panels (a)-(d) show the one-dimensional correlation map as a function of east-west shift speed and latitude (dark is high- and white is low-correlation) -- the month/day/year of the measurement is indicated above each panel. The uncertainty in the zonal wind speed arising from the navigational uncertainty is smaller than the spread in the correlation distribution ($\pm$2.1~m~s$^{-1}$ on May~7, 2008; $\pm$3.8~m~s$^{-1}$ on September~7, 2010; $\pm$1.7~m~s$^{-1}$ on January~11, 2011; and $\pm$4.2~m~s$^{-1}$ on August~5, 2011).

\pagebreak
\begin{center}
\includegraphics[width=3.in]{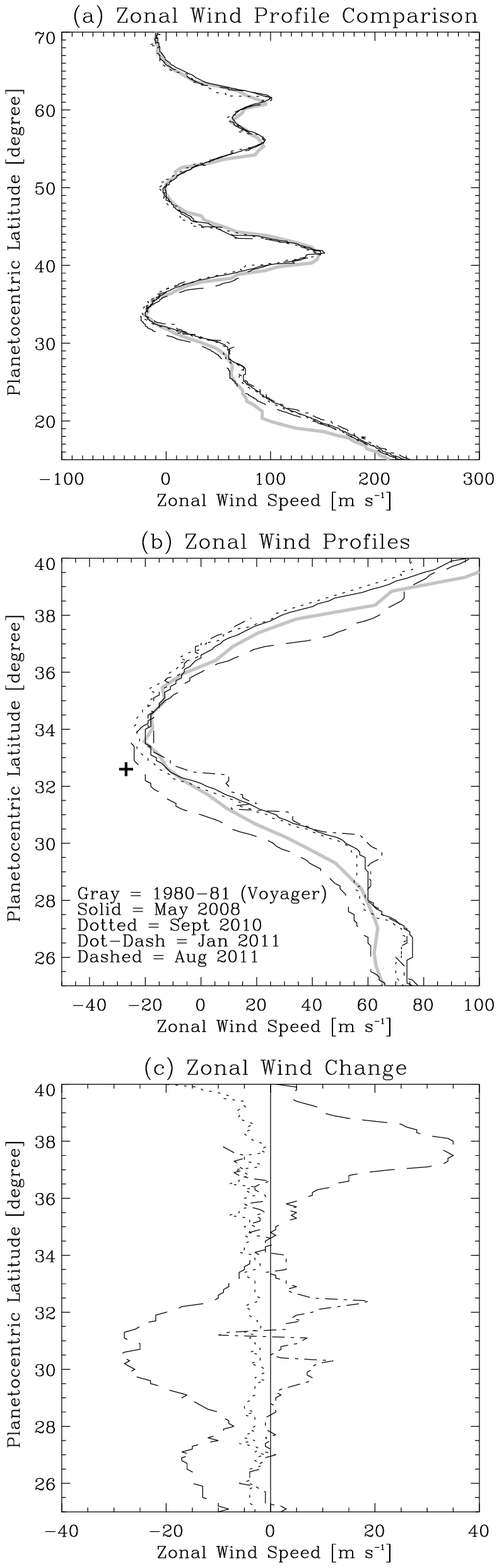}
\end{center}

\noindent \textbf{Figure 15: } 
Comparison of the zonal wind measurements before, during and after the storm's convective activities. All images analyzed here are captured using the CB2 filter. Panel (a) shows the maximum correlation wind profiles for May~7, 2008 (solid), September~7, 2010 (dotted), January~11, 2011 (dash-dot) and August~5, 2011 (dashed). The gray line is the zonal mean wind measurement by Sanchez-Lavega et al. (2000). Panel (b) is the same except that it enlarges the latitudes affected by the storm to highlight the wind speed change. The plus (+) symbol marks the latitude and propagation speed of the storm's head. Panel (c) shows the wind change compared to the May~7, 2008 measurement, illustrating that the measurement after the storm on August~5, 2011 shows a faster wind to the north of the storm and a slower wind to the south.

\pagebreak
\begin{center}
\includegraphics[angle=90, width=5.5in]{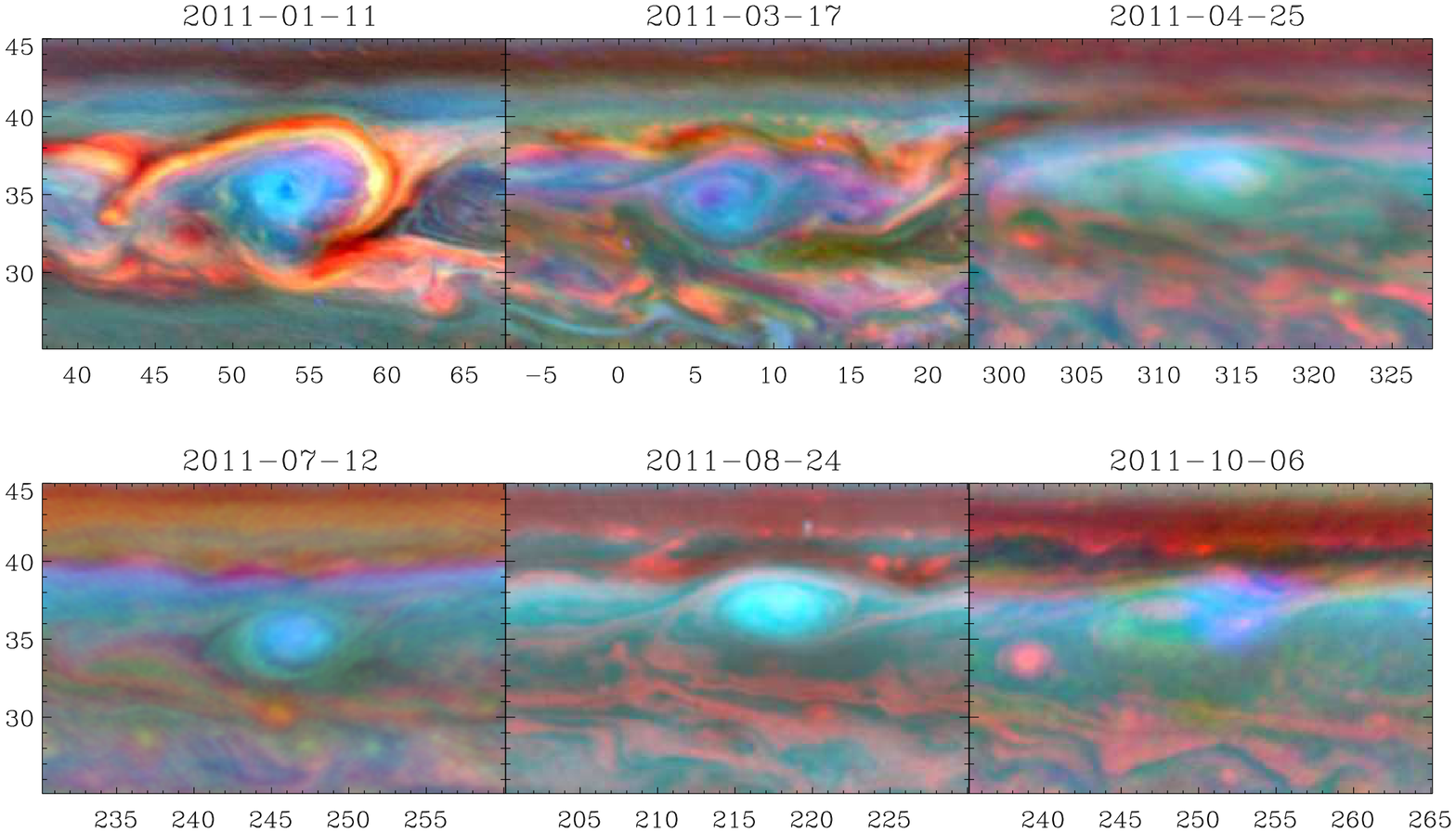}
\end{center}

\noindent \textbf{Figure 16: }
Temporal evolution of AV. The color scheme is the same as in Fig.~2. Although the AV's size initially rivaled that of Jupiter's Oval BA, it continuously shrinks after its formation. It's center also drifts northward as its size decreases.

\pagebreak
\begin{center}
\includegraphics[width=6in]{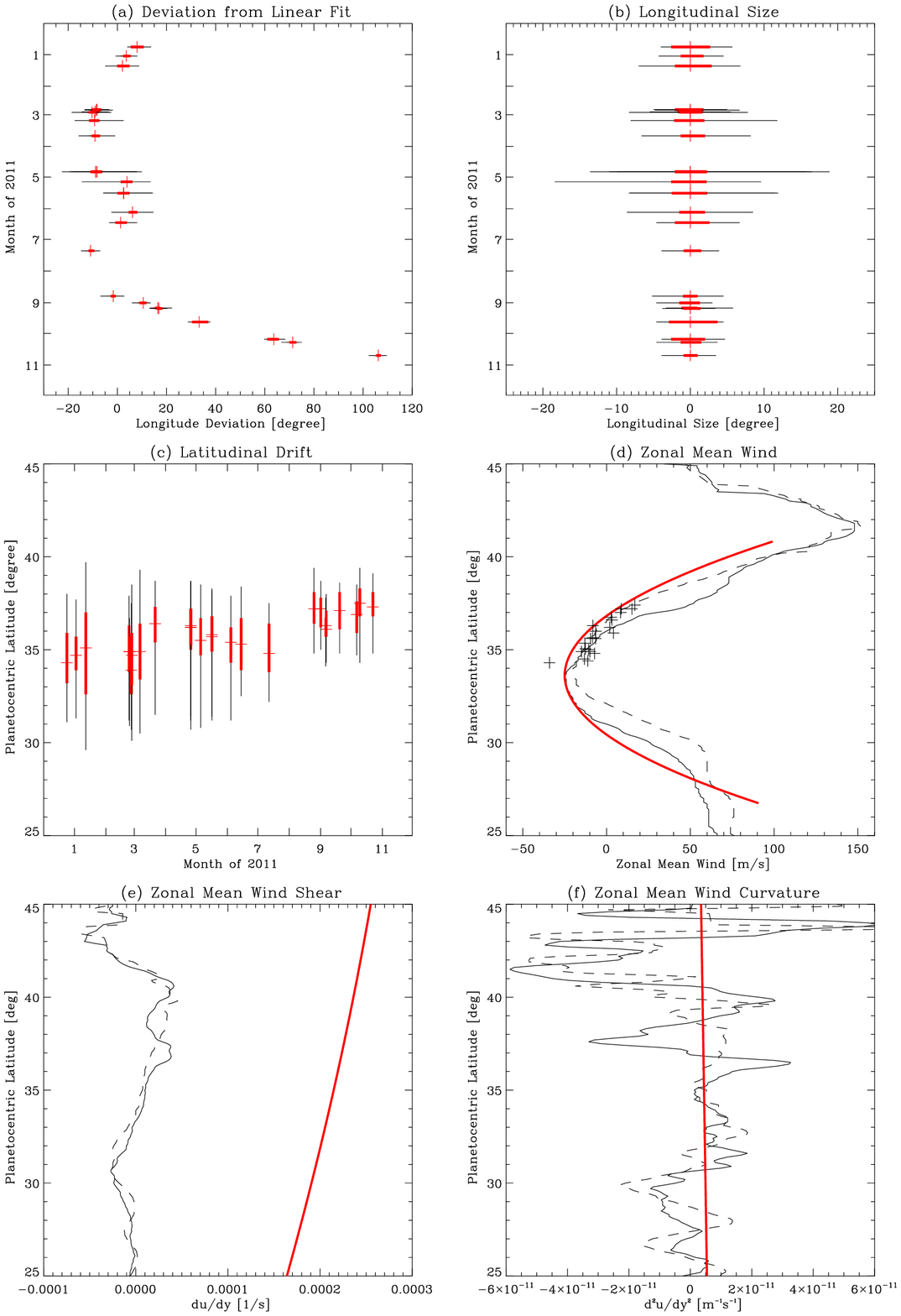}
\end{center}

\noindent \textbf{Figure 17: } 
(a) The deviation of the AV's center longitude from Eq.~3 as a function of time. The red crosses (+) denote the AV's center, the red lines denote the size of AV's round core in CB2, and the solid black line denotes the size of the swirly bright cloud around the AV. (b) Same as (a), except that the AV is centered to illustrate the longitudinal size as a function of time. (c) The center latitude of the AV as a function of time. In panels d-f, the dashed and solid lines represent the wind profiles on May 7, 2008, and August~5, 2011, respectively. (d) Zonal mean wind profiles as functions of latitude. The red line depicts the maximum curvature of the wind allowed around 33\degree N latitude by the Rayleigh-Kuo stability criterion. The plus marks (+) illustrate the zonal drift speeds and latitude of the AV at different times. (e) Zonal mean wind shear $d\bar{u}/dy$ compared to the Coriolis parameter $f$ (red). (f) Zonal mean wind curvatures $d^{2}\bar{u}/dy^{2}$ and $\beta$ (red).

\end{noindent}
\end{document}